# Drift of sensitive direction of Hall-effect devices in (100)-silicon caused by mechanical shear stress

Authors: Udo Ausserlechner[1], Michael Holliber[1], Benjamin Kollmitzer[1], Richard Heinz[2]

Affiliations: [1] are with Infineon Technologies Austria AG, Siemensstrasse 2, A-9500 Villach, Austria.

[2] is with Infineon Technologies AG, Am Campeon 1-15, 85579 Neubiberg, Germany.

Corresponding author: Udo Ausserlechner, Siemensstrasse 2, A-9500 Villach, Austria.
Email: udo.ausserlechner@infineon.com

Abstract:

The output signal of classical symmetrical Hall plates is an odd function of the magnetic field component acting perpendicular to the plate. At weak magnetic field the Hall plate output is linearly proportional to the perpendicular magnetic field. Magnetic field components parallel to the plate may also contribute to the output signal via the planar Hall effect. It leads to even order terms of the in-plane magnetic field in the output signal. At moderate magnetic field the planar Hall effect adds to the output signal a term proportional to the square of the in-plane magnetic field. This paper reports on linear terms of the in-plane magnetic field component to the output signal of Hall-plates, when they are subjected to mechanical shear stress. The effect is small for Hall plates but large for Vertical Hall devices in (100)-silicon. It is fully described by piezo-resistance and piezo-Hall tensors. We present results of numerical simulations and measurements. Thin devices are less affected than thick devices. If magnetic angle sensors are made of Vertical Hall devices, in-plane shear stress leads to a small orthogonality error which – in contrast to the planar Hall effect – cannot be cancelled out by spinning current schemes. We propose a compensation circuit to eliminate this shear-stress induced orthogonality error.

1 Introduction:

In silicon technology a typical Hall plate is made of a shallow tub diffused into the surface of a semiconductor wafer with four contacts arranged on the perimeter of the tub. The tub depth is 10 to 100 times smaller than its lateral size. The current streamlines are parallel to the ($x,y$) wafer surface. Therefore the device responds to the magnetic field along the $z$-direction. This at least is the simple truth of introductory text books. However, in practice there are a few cases when the output signal of Hall plates also responds to lateral magnetic field.

If the n-doped tub of the Hall plate is isolated against the p-substrate by a reverse biased pn-junction the junction field effect occurs (see Figure 1) [1]. It means that the depletion layer of the pn-junction grows versus reverse bias voltage. Therefore, it is thicker at the positive supply contact than at the negative one. Thus the Hall plate is thinner at the positive supply contact – it has a tapered shape. Consequently the current streamlines are not exactly parallel to the wafer surface, but on average

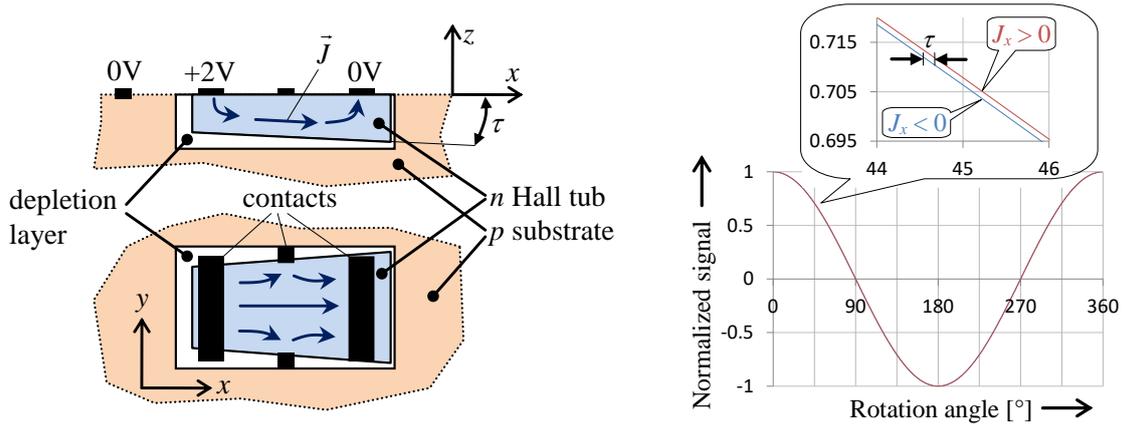

FIGURE 1: Brick-wall model of the junction field effect in Hall plates. The *n*-doped Hall tub is immersed into *p*-substrate. Substrate is tied to 0 V. The supply contacts of the Hall plate are tied to +2 V and 0 V. Then a depletion layer establishes along the reverse biased pn-junction between Hall tub and substrate. The thickness of the depletion layer is larger at the positive supply potential. Therefore the bottom surface of the conductive Hall tub is tilted by an angle $\tau$ against the surface of the substrate. The current flow lines are tilted by $\tau/2$ on average. As a consequence the Hall plate not only responds to vertical magnetic field in *z*-direction, but it also slightly responds to horizontal magnetic field along *x*-direction. If the supply contacts are swapped, the sign of the tilt angle changes. If the Hall plate is rotated in a static magnetic field, the output voltages for both supply polarities are sine waves with a phase shift $\tau$. [Figure in black & white]

they are tilted against the wafer surface by half the taper angle, *i.e.* by $\tau/2$. At weak magnetic field the output signal of the device is proportional to $B_z \cos(\tau/2) - B_x \sin(\tau/2)$, if the supply contacts are aligned with the *x*-axis. When such a Hall plate is rotated around its *y*-axis while a static homogeneous magnetic field perpendicular to the *y*-axis is applied to the chip the output signal is a sine versus rotation angle. If the polarity of the Hall supply current is inverted, one gets another sine output signal. Both sine curves are phase shifted by an angle equal to $\tau$, which is in the order of 0.1°. This was observed in an experiment [2]. If the Hall plate is operated in a spinning current scheme [3] where supply current flows between various contact pairs and with both polarities the effect of the tapering is annihilated. Then the Hall plate has a very high directional selectivity irrespective of the junction field effect – the spinning current Hall plate responds only to $B_z$ field. This can be used to measure the warpage of a die inside a plastic encapsulated package with sub-micrometer resolution [4].

It should be noted that Hall plates usually have top plates, which shield the Hall region from external electric fields to prevent lifetime drift [5]. If interconnect layers (metal or poly-silicon) are used for the top plate, Figure 1 applies. However, quite frequently the top plate is a shallow p-doped tub at the top surface of the substrate. If such a top plate is grounded, the Hall region is bounded by top and bottom depletion layers. If both depletion layers are identical, this avoids net tilting of the current streamlines and the effect of Figure 1 does not apply. In CMOS technologies the top plate is usually made from p$^+$S/D (source-drain) diffusion, which has orders of magnitude higher doping than the substrate, and therefore the depletion layers at the bottom and at the top of the Hall region are hardly ever identical.

Vertical Hall devices are elongated tubs with lengths greater than widths. One tries to use deepest possible tubs, but due to technological limitations in silicon CMOS processes the depth is usually only 3-5 µm and therefore comparable to the width. On the top surface a couple of contacts is placed (three, four, five, sometimes even more). The contacts are spaced apart along the length direction, they extend over the major part of the width, and they are parallel. If current is sent through some contacts the potentials on the other contacts change with magnetic field along the width direction. A widespread type of Vertical Hall device has five contacts like in Figure 2 [6]. We will refer to it throughout this paper. In such devices the following two phenomena are known, which change the direction of maximal sensitivity to magnetic field.

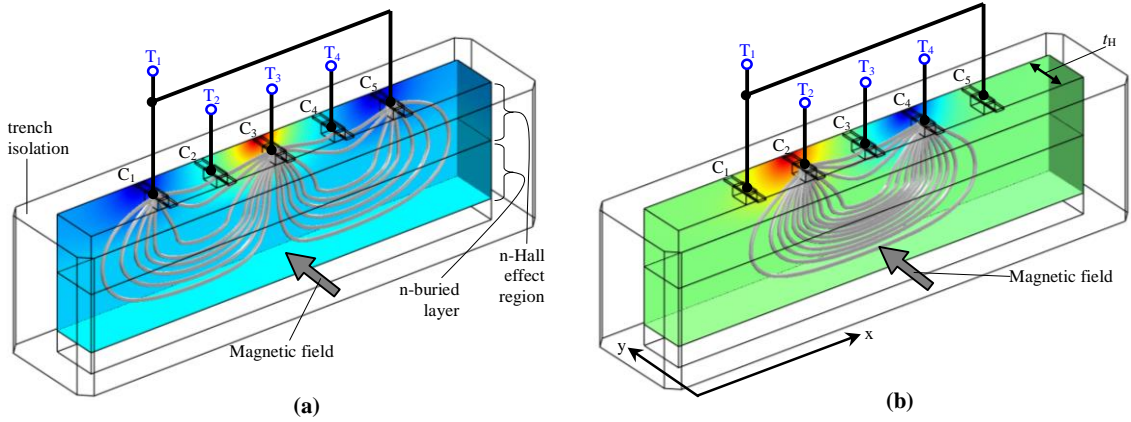

FIGURE 2: A Vertical Hall device with five top contacts and four terminals in BiCMOS technology with deep trench isolation. The contacts comprise shallow $n^+$S/D-diffusions and deeper $n$CMOS-wells. Figure 2(a) shows an operating mode where current flows from the centre contact to the outer ones and the output signal is tapped between terminals $T_2$ and $T_4$. In Figure 2(b) supply and output terminals are swapped. For equal current the output voltage is identical in both cases (provided that linear electrostatic theory applies). At zero shear stress the output signal responds only to the magnetic field in $y$-direction. The thickness $t_H$ of the device extends also in $y$-direction. Due to technological limitations the bottom of the Hall effect region is shorted by a highly conductive $n$-buried layer at floating potential. (reprint from [6]) [Figure in black & white]

At zero mechanical stress and for moderate magnetic flux densities below 2 T in silicon the Hall effect is described by $\vec{E} = \vec{\rho} \cdot \vec{J}$ with ([7])

$$\vec{\rho} = \begin{pmatrix} \rho + \rho\mu_H^2 C(B_y^2 + B_z^2) & -C_H B_z - \rho\mu_H^2 C B_x B_y & C_H B_y - \rho\mu_H^2 C B_x B_z \\ C_H B_z - \rho\mu_H^2 C B_x B_y & \rho + \rho\mu_H^2 C(B_x^2 + B_z^2) & -C_H B_x - \rho\mu_H^2 C B_y B_z \\ -C_H B_y - \rho\mu_H^2 C B_x B_z & C_H B_x - \rho\mu_H^2 C B_y B_z & \rho + \rho\mu_H^2 C(B_x^2 + B_y^2) \end{pmatrix} \qquad (1)$$

Here $\rho$ is the isotropic resistivity at zero magnetic field and zero stress, $\mu_H$ is the Hall mobility, $C_H$ is the Hall coefficient with $|C_H| = \rho|\mu_H|$, $C$ is a dimensionless constant that describes the planar Hall effect, and $B_x, B_y, B_z$ are the components of the flux density. In the off-diagonal elements of the resistivity tensor the dominant term is the Hall coefficient times a component of the flux density – it describes the transverse Hall effect. Yet, the off-diagonal terms also comprise a term proportional to $C$ and to the product of the two other components of the flux density field. Obviously there is a crosstalk of the components $B_i B_j$ into the $B_k$-component ($i \neq j \neq k \neq i \wedge i, j, k \in \{x, y, z\}$). The largest crosstalk occurs if a field of magnitude $B$ points along the symmetry line between coordinates $i$ and $j$ (i.e. $B_i = B_j = B/\sqrt{2}$). Then the ratio of transverse over planar Hall electric fields is $\rho\mu_H^2 C B_i B_j / (|C_H| B_k) \rightarrow |\mu_H| CB/2$. For a Hall plate with CMOS n-well doping of $3.5 \times 10^{16}/\text{cm}^3$ and $|\mu_H| = 0.118/\text{T}$ [7] finds a crosstalk of 8.1% at 2 T. This corresponds to $C = 0.686$. In common automotive sensor systems the flux density does not exceed 0.1 T, which gives only 0.4% crosstalk or 0.23° angle error. Moreover, the effect is smaller in Vertical Hall devices than in Hall plates [7]. The planar Hall effect is proportional to the product of the two components $B_i B_j$, and therefore its signal appears on the second harmonic (*i.e.* with 180° periodicity) when the device is rotated in the (*i,j*)-plane while being subjected to a static magnetic field. Thus, in experiments it can be easily discriminated against effects on the first harmonic like the shear stress induced crosstalk that we investigate in the main body of this paper. In (1) the planar Hall effect adds terms with even symmetry to the resistivity tensor. According to linear network theory, the passivity of the device (*i.e.* it dissipates energy) and the reciprocity of its resistivity tensor allow for its representation by a resistor network [8]. Thereby the output signal of the planar Hall effect is reflected by an asymmetry of this network. However, any kind of asymmetry in the equivalent resistor circuit is cancelled out by spinning current schemes [9]. This was also proven experimentally for Hall plates and Vertical Hall devices by [10].

At large magnetic field (above 0.2 T in n-doped (100)-silicon) the anisotropy of the semiconductor crystal may change the direction of maximal sensitivity of Vertical Hall devices. A strong shift of 4° at 2 T was found in measurements on devices of the type shown in Figure 2(a) in [11], if they are aligned at 15° or 30° to the [110] axes. If the devices are aligned to [110] or [100] axes, no angular shift is found at high magnetic field. Thus, the effect concerns only unconventional device orientations. The root cause of this effect is anisotropy of the transport process caused by large magnetic field. In the following we will study a similar effect, yet in our case the anisotropy is caused by mechanical stress at weak magnetic field.

It is known that a large and poorly controlled mechanical stress is exerted by plastic encapsulated packages on semiconductor dies. This has a significant impact on the energy band structure of indirect semiconductors such as silicon and germanium. Amongst all electronic parameters of circuit devices, mobility and Hall coefficient are most prone to stress. This affects resistors, MOS and bipolar transistors, and Hall devices. In (100)-silicon normal stress components in the chip surface ($\sigma_{xx}, \sigma_{yy}$) and orthogonal to it ($\sigma_{zz}$) change the magnetic sensitivity of Hall plates and Vertical Hall devices at constant input current. This is mainly caused by the piezo-Hall effect, which acts on the Hall coefficient [12]. Yet a small part may also come from piezo-resistance effect, which diverts the current flow lines thereby changing the Hall geometry factor (cf. section 5 of [13]). Stress is also a major source of offset error (zero point error) of Hall plates and to a lesser extent of Vertical Hall devices [13,13b]. Stress related offset can be explained by piezoresistance. The effect is intentionally used to measure certain stress components in pseudo-Hall or Kanda devices ([14], [15], see also [16]). In Hall devices it is cancelled out by spinning current schemes. So far there has not been any discussion of crosstalk of orthogonal magnetic field components in the output signals of Hall effect devices caused by mechanical stress. Only a short appendix on crosstalk by chip warping is given in a paper of stray field robust magnetic angle sensors with Hall plates [17]. Orthogonality errors of Hall based magnetic angle sensors are briefly mentioned in [18, 19] as being caused by oblique implantation angles, inhomogeneous doping and mask imperfection – however not by mechanical stress. In the following we will show that shear stress may lead to a notable crosstalk of orthogonal magnetic field components in Vertical Hall devices.

## 2 Theory of piezo-resistance and piezo-Hall in silicon Hall-effect devices

### 2.1 The conductivity matrix at mechanical stress and magnetic field in 3D for cubic crystals

Silicon wafers are made of single crystals in a diamond lattice which belongs to the group of cubic Bravais lattices. The crystal symmetry is reflected in the symmetry of the tensors that describe anisotropic transport phenomena in these materials. At zero mechanical stress and zero magnetic field the resistivity $\rho$ is isotropic in cubic crystals.

$$\vec{E} = \rho \cdot \vec{J} \tag{2}$$

$\vec{E}$ is the electric field and $\vec{J}$ is the charge current density. $\vec{E}$ and $\vec{J}$ are parallel.

The magnetic field introduces anisotropy in the charge transport – this is given by the Hall coefficient $C_H$ in (1). For small magnetic field we can neglect the planar Hall effect and rewrite (1) like this

$$\vec{E} = \vec{\tilde{\rho}} \cdot \vec{J} \tag{3a}$$

with the anti-symmetric second rank tensor $\vec{\tilde{\rho}}$.

$$\vec{\tilde{\rho}} = \begin{pmatrix} \rho & -C_H B_z & C_H B_y \\ C_H B_z & \rho & -C_H B_x \\ -C_H B_y & C_H B_x & \rho \end{pmatrix} \tag{3b}$$

$\vec{E}$ and $\vec{J}$ are not parallel.

If mechanical stress acts on the silicon at zero magnetic field, the resistivity becomes anisotropic according to the crystal symmetry.

$$\vec{\tilde{\rho}}^{crystal} = \rho \left\{ \begin{pmatrix} 1 & 0 & 0 \\ 0 & 1 & 0 \\ 0 & 0 & 1 \end{pmatrix} + \begin{pmatrix} \pi_{11} & \pi_{12} & \pi_{12} & 0 & 0 & 0 \\ \pi_{12} & \pi_{11} & \pi_{12} & 0 & 0 & 0 \\ \pi_{12} & \pi_{12} & \pi_{11} & 0 & 0 & 0 \\ 0 & 0 & 0 & \pi_{44} & 0 & 0 \\ 0 & 0 & 0 & 0 & \pi_{44} & 0 \\ 0 & 0 & 0 & 0 & 0 & \pi_{44} \end{pmatrix} \cdot \begin{pmatrix} \sigma_{xx}^{crystal} \\ \sigma_{yy}^{crystal} \\ \sigma_{zz}^{crystal} \\ \sigma_{yz}^{crystal} \\ \sigma_{xz}^{crystal} \\ \sigma_{xy}^{crystal} \end{pmatrix} \right\} \tag{4}$$

In (4) we used the reference frame of the crystal with $x^{crystal}, y^{crystal}, z^{crystal}$-axes identical to the $[1,0,0], [0,1,0], [0,0,1]$-directions of the single crystal. We also used the 6x1-vector representation for the second rank stress tensor $\vec{\tilde{\sigma}}$

$$\begin{pmatrix} \sigma_{xx} & \sigma_{xy} & \sigma_{xz} \\ \sigma_{xy} & \sigma_{yy} & \sigma_{yz} \\ \sigma_{xz} & \sigma_{yz} & \sigma_{zz} \end{pmatrix} = \begin{pmatrix} \sigma_{xx} \\ \sigma_{yy} \\ \sigma_{zz} \\ \sigma_{yz} \\ \sigma_{xz} \\ \sigma_{xy} \end{pmatrix} \tag{5}$$

and the 6x6-matrix representation for the fourth rank piezo-resistance tensor $\vec{\tilde{\pi}}$ [20, 21]. In the piezo-resistance tensor (4) several elements vanish and others are identical due to the symmetry of the crystal. Finally one needs only three numbers $\pi_{11}, \pi_{12}, \pi_{44}$. The piezo-resistance tensor describes the change in resistivity caused by mechanical stress. $\vec{\tilde{\rho}}$ has even symmetry.

If both mechanical stress and magnetic field are present the charge transport is governed by [12, 22, 23][1]

$$\vec{E}^{crystal} = \vec{\vec{\rho}}^{crystal} \cdot \vec{J}^{crystal} + \left(\vec{\vec{C}}_H^{crystal} \cdot \vec{B}^{crystal}\right) \times \vec{J}^{crystal}. \tag{6}$$

In (6) the Hall coefficient $C_H$ is a second rank tensor.

$$\vec{\vec{C}}_H^{crystal} = C_H \left\{ \begin{pmatrix} 1 & 0 & 0 \\ 0 & 1 & 0 \\ 0 & 0 & 1 \end{pmatrix} + \begin{pmatrix} P_{11} & P_{12} & P_{12} & 0 & 0 & 0 \\ P_{12} & P_{11} & P_{12} & 0 & 0 & 0 \\ P_{12} & P_{12} & P_{11} & 0 & 0 & 0 \\ 0 & 0 & 0 & P_{44} & 0 & 0 \\ 0 & 0 & 0 & 0 & P_{44} & 0 \\ 0 & 0 & 0 & 0 & 0 & P_{44} \end{pmatrix} \cdot \begin{pmatrix} \sigma_{xx}^{crystal} \\ \sigma_{yy}^{crystal} \\ \sigma_{zz}^{crystal} \\ \sigma_{yz}^{crystal} \\ \sigma_{xz}^{crystal} \\ \sigma_{xy}^{crystal} \end{pmatrix} \right\} \tag{7}$$

$\vec{\vec{P}}$ is the fourth rank piezo-Hall tensor with the same symmetry as the piezo-resistance tensor and with only three non-vanishing numbers $P_{11}, P_{12}, P_{44}$. The piezo-Hall tensor describes the change of Hall coefficient caused by mechanical stress.

Inserting (4) and (7) into (6) gives the relation $\vec{E}^{crystal} = \vec{\vec{\rho}}_{effective}^{crystal} \cdot \vec{J}^{crystal}$. According to Onsager $\vec{\vec{\rho}}_{effective}^{crystal}$ exhibits reverse magnetic field reciprocity (RMFR), *i.e.* columns and rows are swapped if the polarity of the magnetic field is inverted [25]. $\vec{\vec{\rho}}_{effective}^{crystal}$ is a function of the magnetic field and the mechanical stress – all of them measured in the reference frame of the single crystal.

In silicon CMOS technologies the wafers are (100)-planes and the x- and y-axes of the chip are aligned along $\langle 110 \rangle$-crystal directions as shown in Fig. 3 (see also [26]). Thus, all vector and tensor quantities have to be transformed into the chip reference frame [21]. The result is $\vec{E}^{chip} = \vec{\vec{\rho}}_{effective}^{chip} \cdot \vec{J}^{chip}$. The components of the resistivity tensor are

$$\rho_{effective,xx}^{chip} = \rho\left(1 + \frac{\pi_{11} + \pi_{12} + \pi_{44}}{2}\sigma_{xx}^{chip} + \frac{\pi_{11} + \pi_{12} - \pi_{44}}{2}\sigma_{yy}^{chip} + \pi_{12}\sigma_{zz}^{chip}\right) \tag{8a}$$

$$\rho_{effective,yy}^{chip} = \rho\left(1 + \frac{\pi_{11} + \pi_{12} - \pi_{44}}{2}\sigma_{xx}^{chip} + \frac{\pi_{11} + \pi_{12} + \pi_{44}}{2}\sigma_{yy}^{chip} + \pi_{12}\sigma_{zz}^{chip}\right) \tag{8b}$$

$$\rho_{effective,zz}^{chip} = \rho\left(1 + \pi_{12}\sigma_{xx}^{chip} + \pi_{12}\sigma_{yy}^{chip} + \pi_{11}\sigma_{zz}^{chip}\right) \tag{8c}$$

$$\rho_{effective,yz}^{chip} = \rho\pi_{44}\sigma_{yz}^{chip} - C_H \begin{pmatrix} 1 + \frac{P_{11} + P_{12} + P_{44}}{2}\sigma_{xx}^{chip} + \frac{P_{11} + P_{12} - P_{44}}{2}\sigma_{yy}^{chip} + P_{12}\sigma_{zz}^{chip} \\ (P_{11} - P_{12})\sigma_{xy}^{chip} \\ P_{44}\sigma_{xz}^{chip} \end{pmatrix} \cdot \begin{pmatrix} B_x^{chip} \\ B_y^{chip} \\ B_z^{chip} \end{pmatrix} \tag{8d}$$

---

[1] Hälg and Keyes seem to have used the wrong sign with the Hall constant – Price has the right sign (which also agrees with Seeger [24]). The problem with the sign may originate from different definitions of the Hall coefficient: usually, it is negative for electrons and positive for holes.

$$\rho^{chip}_{effective,xz} = \rho\pi_{44}\sigma^{chip}_{xz} + C_H \left( 1 + \frac{P_{11}+P_{12}-P_{44}}{2}\sigma^{chip}_{xx} + \frac{P_{11}+P_{12}+P_{44}}{2}\sigma^{chip}_{yy} + P_{12}\sigma^{chip}_{zz} \quad (P_{11}-P_{12})\sigma^{chip}_{xy} \atop P_{44}\sigma^{chip}_{yz} \right) \cdot \begin{pmatrix} B^{chip}_x \\ B^{chip}_y \\ B^{chip}_z \end{pmatrix} \quad (8e)$$

$$\rho^{chip}_{effective,xy} = -\rho(\pi_{12}-\pi_{11})\sigma^{chip}_{xy} - C_H \begin{pmatrix} P_{44}\sigma^{chip}_{xz} \\ P_{44}\sigma^{chip}_{yz} \\ 1 + P_{12}(\sigma^{chip}_{xx}+\sigma^{chip}_{yy}) + P_{11}\sigma^{chip}_{zz} \end{pmatrix} \cdot \begin{pmatrix} B^{chip}_x \\ B^{chip}_y \\ B^{chip}_z \end{pmatrix} \quad (8f)$$

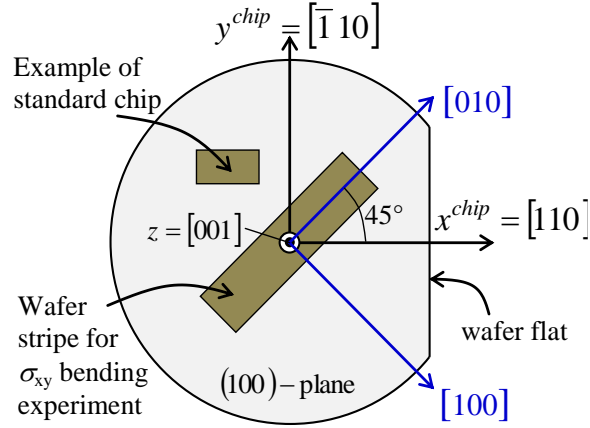

FIGURE 3: Reference frame $\langle 100 \rangle$ of the single crystal and chip reference frame $\Sigma = (x, y, z)$ for a (100)-silicon wafer. For commercial products standard chips are cut out of the wafer with their edges parallel to $[110]$ and $[\bar{1}10]$ directions. For our bending experiment in section 4 wafer stripes were cut out with the long edges parallel to the $[010]$ crystal direction. [Figure in black & white]

The conductivity is the inverse of the resistivity $\vec{\kappa}^{chip}_{effective} = (\vec{\rho}^{chip}_{effective})^{-1}$ (see Appendix B). The other components of the resistivity and conductivity tensors can be obtained by the RMFR, *e.g.* one gets $\kappa^{chip}_{effective,yx}$ simply by changing the polarity of all magnetic field components in (B6). Equations (8a-f) also work for large magnetic field. Equations (B1-B6) are approximations for small magnetic field.

## 2.2 Shear stress on Vertical Hall devices

In this section we refer to the chip coordinate system, whereby we mostly discard the superscript "chip" for the sake of brevity. We consider a device for $B_y$-detection. Its tub extends in *x*- and *y*-directions but the contacts are lined up in *x*-direction, whereby each contact extends mainly along *y*-direction. Fig. 2 shows an example of such a device. Moreover, we consider only the dominant in-plane stress components $\sigma_{xx}, \sigma_{yy}, \sigma_{xy}$ along the chip surface.

### 2.2.1 The slim Vertical Hall device

If the width of the tub along *y*-direction is small, the current is confined to the (*xz*)-plane. We compute the *y*-component of the current density $J^{chip}_y$ from $\vec{J}^{chip} = \vec{\kappa}^{chip}_{effective} \cdot \vec{E}^{chip}$ with (B2), (B4), (B6). Then we set it equal to zero and solve for $E^{chip}_y$. We insert this into the other equations for $J^{chip}_x, J^{chip}_z$. After linearization for small magnetic field and small stress we get

$$\frac{\vec{\rho}^{(xz,\text{thin})}}{\rho} = \begin{pmatrix} 1 + (\pi_{11} + \pi_{12})\frac{\sigma_{xx} + \sigma_{yy}}{2} + \pi_{44}\frac{\sigma_{xx} - \sigma_{yy}}{2} & \frac{C_H}{\rho}\left(uB_y + (P_{11} - P_{12})\sigma_{xy}B_x\right) \\ -\frac{C_H}{\rho}\left(uB_y + (P_{11} - P_{12})\sigma_{xy}B_x\right) & 1 + \pi_{12}(\sigma_{xx} + \sigma_{yy}) \end{pmatrix} \quad (9a)$$

$$u = 1 + (P_{11} + P_{12})\frac{\sigma_{xx} + \sigma_{yy}}{2} - P_{44}\frac{\sigma_{xx} - \sigma_{yy}}{2} \quad (9b)$$

Equations (9a,b) mean that at zero stress the Hall electric field is proportional to $C_H B_y$, but under the action of in-plane shear stress the term $C_H(P_{11} - P_{12})\sigma_{xy}B_x$ is added. Unfortunately, the piezo-Hall coefficient for low n-doped silicon is large $P_{11} - P_{12} = -137\%/\text{GPa}$. The normal stress $\sigma_{xx}, \sigma_{yy}$ scales the elements in the main diagonal and the $B_y$-field, but it does not lead to crosstalk of $B_x$ and $B_y$.

In a real device with isolating boundary and with contacts of finite size the output voltage is given by

$$V_{out} = C_H t_H^{-1} G_{H0} I_{in} B_y + V_{off} \quad (10a)$$

where $t_H$ is the thickness of the device (*i.e.* the width in y-direction), $I_{in}$ is the current through the device, and $V_{off}$ is the raw offset voltage at zero applied magnetic field. $0 \leq G_{H0} \leq 1$ is the weak field Hall geometry factor which describes the loss in signal due to short circuiting effects of the contacts. In the presence of shear stress it holds

$$V_{out} = C_H t_H^{-1} G_{H0} I_{in} \left(B_y + (P_{11} - P_{12})\sigma_{xy}B_x\right) + V_{off}. \quad (10b)$$

The same Hall geometry factor applies because we simply replace $B_y \rightarrow B_y + (P_{11} - P_{12})\sigma_{xy}B_x$ in both off-diagonal elements of the resistivity tensor $\vec{\rho}^{(xz,\text{thin})}$. Therefore we cannot modify the crosstalk between $B_x$ and $B_y$ by geometrical changes in the (x,z)-plane cross-section of the device. Note that in general $G_{H0}$ and $V_{off}$ depend on $\sigma_{xx}$ and $\sigma_{yy}$ (as it is explained in [13]).

### 2.2.2 The thick Vertical Hall device

For very large width of the tub along y-direction the y-component of the electric field $E_y^{chip}$ vanishes. This is due to the contacts, which extend along the entire width, thereby shorting electric field along the width. Therefore, we start with the resistivity tensor, compute $E_y^{chip}$ and set it equal to zero. We solve this for $J_y^{chip}$ and insert it back into the other equations for $E_x^{chip}, E_z^{chip}$. After linearization for small magnetic field and small stress, we get the resistivity tensor

$$\frac{\vec{\rho}^{(xz,\text{thick})}}{\rho} =$$

$$= \begin{pmatrix} 1 + (\pi_{11} + \pi_{12})\frac{\sigma_{xx} + \sigma_{yy}}{2} + \pi_{44}\frac{\sigma_{xx} - \sigma_{yy}}{2} & \frac{C_H}{\rho}\left(uB_y + (P_{11} - P_{12} + \pi_{11} - \pi_{12})\sigma_{xy}B_x\right) \\ -\frac{C_H}{\rho}\left(uB_y + (P_{11} - P_{12})\sigma_{xy}B_x\right) & 1 + \pi_{12}(\sigma_{xx} + \sigma_{yy}) \end{pmatrix} \quad (11)$$

In (11) the crosstalk between *x*- and *y*-components of the magnetic field is even larger than for thin devices: $P_{11} - P_{12} + \pi_{11} - \pi_{12} = -293.6 \%/\text{GPa}$. The reason for the stronger effect in thick devices can be seen in the 3D conductivity tensor $\kappa_{effective,xy}^{chip}$ in (B6). There we note a piezo-resistive term $\rho^{-1}(\pi_{12} - \pi_{11})\sigma_{xy}^{chip}$ which pushes a $J_x^{chip}$-current slightly into *y*-direction at zero magnetic field. The small deviation angle is $(\pi_{11} - \pi_{12})\sigma_{xy}^{chip}$. Thus, in thick devices this extra tilt in current due to piezo-resistance adds to the piezo-Hall effect of the thin devices. Conversely, the piezo-Hall coefficients in (9a,b) come from another origin: the charge carriers are scattered into a slightly different direction at mechanical stress. This is described by the anisotropy of the Hall coefficient at mechanical stress (7).

The two cases of thin versus thick Vertical Hall devices are analogous to the states of plain stress versus plain strain in structural mechanics. Consider a thin wafer stripe being long in x-direction, but slim in y-direction ($L_y \ll L_x$). The bending of the stripe by a moment $M_y$ is described by a plain stress model in the (*xz*)-plane with $\sigma_{yy} = 0$. The condition $\sigma_{yy} = 0$ corresponds to $J_y^{chip} = 0$ in the thin Vertical Hall device. Conversely, if the width of the wafer stripe in *y*-direction is not small against the length, the moment $M_y$ gives rise to cylindrical bending, which is a plain strain problem in the (*xz*)-plane with $\varepsilon_{yy} = 0$. For orthotropic materials it holds $E\varepsilon_{yy} = \sigma_{yy} - \nu\sigma_{xx} - \nu\sigma_{zz}$ [27] and this gives $\sigma_{yy} = \nu(\sigma_{xx} + \sigma_{zz})$ ($\nu$ is the Poisson ratio). The stress in width direction in wide stripes during cylindrical bending corresponds to the current flow component in width direction in the thick Vertical Hall devices caused by piezo-resistance. For wide stripes it holds $\varepsilon_{yy} = 0$, and for thick Vertical Hall devices it holds $E_y^{chip} = 0$. If the wafer stripe is neither slim nor very wide it holds $\sigma_{yy} = \eta\nu(\sigma_{xx} + \sigma_{zz})$ with $0 \leq \eta \leq 1$. Analogously, we can write for Vertical Hall devices of arbitrary width $t_H$

$$V_{out} = C_H t_H^{-1} G_{H0} I_{in} \left( B_y + (P_{11} - P_{12} + \eta(\pi_{11} - \pi_{12}))\sigma_{xy} B_x \right) + V_{off}. \tag{12}$$

with $0 \leq \eta \leq 1$ and $\eta$ rising monotonously with $t_H$. In (12) one might wonder why for arbitrary $\eta$ $G_{H0}$ should not depend on shear stress. After all shear stress diverts the current streamlines and this may well affect $G_{H0}$. In [13] it was shown that shear stress (or in general any in-plane biaxial stress) changes $G_{H0}$ for Hall plates unless they have 90° symmetry. In the case of Vertical Hall devices in-plane shear stress $\sigma_{xy}$ tilts the current streamlines out of the (*x,z*) cross-section plane. This tilting is inhibited by the finite width of the device. Infinitely wide devices have largest tilting. For them $G_{H0}$ does not depend on $\sigma_{xy}$ because in (11) $\sigma_{xy}$ shows up only in the off-diagonal elements and there it is not multiplied to By. This is all valid in a small stress approximation – at large stress the simple tilting of current streamlines will cause $dV_{out}/dB_y/I_{in} \propto \cos\arctan((\pi_{11} - \pi_{12})\sigma_{xy})$ with lowest order terms of $\sigma_{xy}^2$.

Note that in (11) the piezo-resistive coefficients $\pi_{11}, \pi_{12}$ occur in the off-diagonal elements with odd symmetry, whereas it is a distinctive property of piezo-resistance at zero magnetic field to exhibit even symmetry in the conductivity and resistivity tensors.

**2.3 Shear stress on Hall plates**

For Hall plates the relevant components in the resistivity and conductivity tensors are $\rho_{effective,xy}^{chip}$ and $\kappa_{effective,xy}^{chip}$. Neglecting higher powers in stress the crosstalk between $B_z$ and $B_x$ is equal to $(\eta\pi_{44} + P_{44})\sigma_{xz}^{chip}$. For low n-doped silicon $P_{44}$ is -6%/GPa [28]. This piezo-Hall coefficient is 23 times smaller than $P_{11} - P_{12}$. Moreover, for Hall plates the out-of-plane shear stress components $\sigma_{xz}^{chip}, \sigma_{yz}^{chip}$ are relevant, and they should be smaller than in-plane shear stress in usual packages. Therefore it is unlikely that package induced mechanical stress alters the sensitive direction of thin silicon Hall plates by more than 0.01°.

## 3 Numerical simulation results

In a first finite element simulation of Figure 4 we modeled a Vertical Hall device with five contacts from the type shown in Figure 2(a). The lightly n-doped Hall tub is a 30.5 μm x 30.5 μm square with 5.5 μm depth. Its bottom surface is a perfectly conductive plane that models the highly n-doped buried layer in a BiCMOS technology. It is floating, *i.e.* no net current is flowing into or out of it. The contacts are made by n-CMOS wells with a size of 1.6 μm x 30.5 μm. They reach into a depth of 1.8 μm. Shallow n⁺S/D diffusions are above the n-CMOS wells. They have the same lateral size and a thickness of 0.2 μm. Their top surfaces are modeled by perfectly conducting ideal contacts. The spacing between the center contacts and their neighbors is 4 μm, the spacing between the outer contacts and their neighbors is 6 μm. In the Hall effect region the specific resistivity was $\rho = 0.016\ \Omega\text{m}$, and the Hall coefficient was $C_H = 0.00209\ \text{m}^3/\text{As}$, which corresponds to a Hall mobility of 1300 cm²/Vs. The conductivity tensor from (B1-B6) was used, whereby $\sigma_{xy} = 100\ \text{MPa}$ and all other stress components vanish. The piezo-coefficients were used only in the Hall well. Their values are $\pi_{11} = -1.022 \times 10^{-9}/\text{Pa}$, $\pi_{12} = 5.44 \times 10^{-10}/\text{Pa}$, $P_{11} = -9.3 \times 10^{-10}/\text{Pa}$, $P_{12} = 4.4 \times 10^{-10}/\text{Pa}$. The resistivity of the n-CMOS wells and n⁺S/D-diffusions were 30 times and 1400 times smaller than in the Hall region. The magnetic flux density $B_x = B_0 \cos\alpha$, $B_y = B_0 \sin\alpha$, $B_z = 0$, with $B_0 = 0.1\ \text{T}$ was applied. The rotation angle $\alpha$ was swept from 0° to 360° in steps of 22.5°. A current of 1 mA was injected into the center contact while the outer contacts were grounded. This led to a supply voltage of 0.457 V.

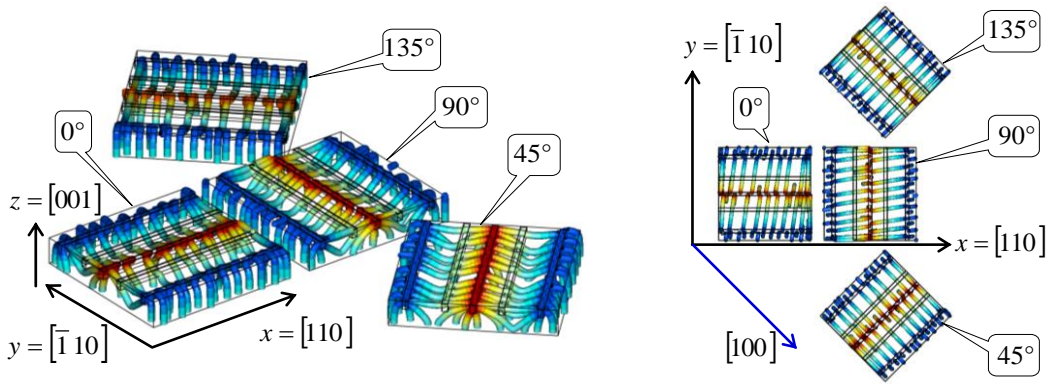

FIGURE 4: A 3D finite element simulation model of Vertical Hall devices with four different orientations 0°, 45°, 90°, 135°. The n-doped Vertical Hall devices are laterally isolated by deep trenches with nitride coating and at their bottom they have a highly conductive n-buried layer which acts like a floating contact. Each device is supplied with 1 mA current at its center contact and grounded at its outer contacts. An in-plane shear stress $\sigma_{xy} = 100\ \text{MPa}$ is assumed. The current streamlines are shown and the color coding denotes the electrostatic potential (red means positive potential, blue means 0 V ground potential). The right figure is a plan view of the (x,y)-plane. There the tilt of the current streamlines due to shear stress is clearly visible for the 0° and 90° devices. The 45° and 135° devices have no such tilt of the current streamlines caused by $\sigma_{xy}$. [Figure in black & white]

The output voltage was tapped between the other contacts on the top surface. Four devices were modeled – the long edges of their contacts were at 0°, 45°, 90°, 135° to the *x*-axis. The discretization mesh had 6.6 million elements. The Laplace equation was solved in 3D space by the commercial

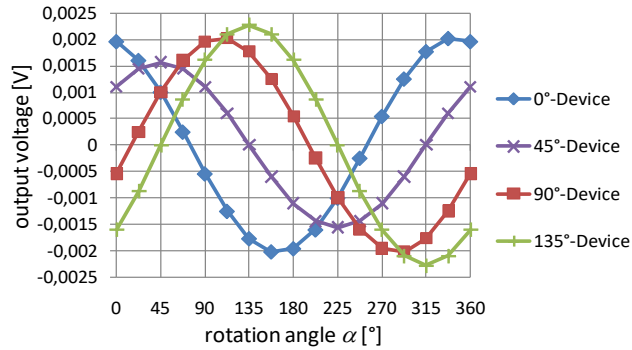

FIGURE 5: Output voltage versus rotation angle as obtained from the finite element simulation of Figure 4 for a rotating magnetic flux density of 100 mT. All curves are perfect sine waves with zero offset. The curves of 0° and 90° devices have identical amplitude, but the output of the 0° device has a positive phase shift and the 90° device has a negative phase shift. The outputs of the 45° and 135° devices have no phase shift, but different amplitudes. [Figure in black & white]

software COMSOL MULTIPHYSICS (application mode emdc). Figure 5 plots the output voltages of these devices versus rotation angle $\alpha$ at a shear stress $\sigma_{xy} = 100$ MPa. All curves are perfect sine waves with no offset and no higher harmonics. The 0° and 90° devices have phase shifted outputs, whereby the 0° output leads the rotation angle by 15.5° while the 90° output lags the rotation angle by 15.5°. The 0° device responds with 19.647 mV/T in x-direction and -5.449 mV/T in y-direction. The 90° device responds with -5.449 mV/T in x-direction and 19.647 mV/T in y-direction. It holds $\arctan 5.449/19.647 = 15.5°$. The amplitudes of 0° and 90° devices are identical. The 45° and 135° devices exhibit no phase shift in their outputs, but they have different amplitudes. This simulation result seems to suggest that 45° and 135° devices are better suited for magnetic angle sensors than 0° and 90° devices, because the former show no phase dependence on in-plane shear stress. However, the phases of 45° and 135° devices depend on $\sigma_{xx} - \sigma_{yy}$ which is an even bigger problem in practice. The volume average of the current density vector throughout the Hall region is rotated by $\arctan\left(\int |J_x| dV / \int |J_y| dV\right) = 8.2°$ against the length direction of the 0° and 90° devices. Thus, the 'average' tilt of current streamlines is less than $\arctan\left((\pi_{11} - \pi_{12})\sigma_{xy}\right) = 8.9°$ because the finite width of the devices constrains it. The shift of 0° and 90° output signals for thin devices would be $\arctan\left((P_{11} - P_{12})\sigma_{xy}\right) = 7.8°$. Thus, it holds $\eta = (\tan 15.5° - \tan 7.8°)/\tan 8.9° = 0.90 < 1$.

We repeated this numerical simulation for devices with different widths, all other parameters were kept constant. Per definition the width direction is perpendicular to the current streamlines at zero mechanical stress and zero magnetic field. Figure 6 plots the computed phase of the sinusoidal output voltage during rotation of the magnetic field versus the width for the 0° Vertical Hall device at a shear stress of 100 MPa. At vanishing stress this phase would also vanish. For a width of 30.5 µm the phase is 15.5° (see above). The phase is ~8° for very thin devices and it increases beyond 16° for fairly thick devices. The theory predicts $\arctan(P_{11} - P_{12})\sigma_{xy} = \arctan(0.137) = 7.8°$ for thin devices and $\arctan(P_{11} - P_{12} + \pi_{11} - \pi_{12})\sigma_{xy} = \arctan(0.2936) = 16.4°$ for thick devices.

Further numerical simulations suggest that the phase shift is identical in both operating modes of Figure 2(a,b). The phase is also independent of the cross sectional geometry of the device. Therefore the spacing and the size of the contacts do not affect the phase as long as the contacts extend over the entire width of the device. Any change in the width direction affects the phase – e.g., if one or more contacts do not extend over the entire width this will also affect the phase versus shear stress.

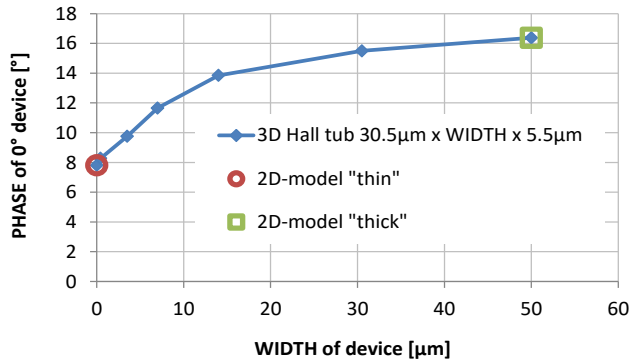

FIGURE 6: Phase of the sinusoidal output signal of the 0° device of Figure 4 during rotation of the magnetic field at $\sigma_{xy}=100$ MPa for varying width of the device (as obtained from a finite element simulation). The limits of thin and thick devices are also indicated. The phase error is 8°/100 MPa for thin devices and it gradually grows versus thickness until it roughly doubles for very thick devices. All contacts extend over the entire width of the devices. The n-doped devices have an n-buried layer at the bottom with a junction isolation to p-substrate and a dielectric trench isolation at the perimeter. [Figure in black & white]

The phase does not even depend on the number of contacts: The phase of Vertical Hall devices with three or four contacts like in [29] and [30] is equally affected by shear stress.

## 4 Experimental investigation

### 4.1 A wafer stripe bending experiment with rotating magnetic field

Here we describe the test fixture that was used to apply in-plane shear stress and 360° rotating in-plane magnetic field to Vertical Hall devices. An overview of the concept behind this measurement is given in Figure 7. A wafer stripe WS was held in a stator ST and bent by a movable piston P, which was coupled via a force sensor FS to a translation stage VT. A the same time a stepper motor RM rotated a permanent magnet PM above the device under test DUT on the top surface of the wafer stripe.

Stripes were cut out of (100)-silicon wafers with their long edges in [010] directions (cf. Figure 3). Bending a stripe in a four-point bending fixture [31] generates in-plane stress $\sigma$ along its long axis, and this stress transforms to the following stress components in the $(x, y, z)$ reference frame of the chip (Mohr's stress circle).

$$\sigma_{xx} = \sigma_{yy} = \sigma_{xy} = \frac{\sigma}{2} \qquad (13)$$

Hence, the bending generates shear stress and normal stress in the $(x, y)$ plane. In this experiment we cannot separate these three in-plane stress components.

The wafer stripes had a length of 40 mm, a width $W$ = 9.9 mm, and a thickness of 752 µm. However, the single crystal silicon is only 739.7 µm thick and on its top side there are 18 thin layers of SiO, SiC, SiN, copper interconnect lines, and Imide coating with a total thickness of 12.3 µm.

The bending force is applied to the wafer stripe via 1.5 mm diameter steel pins. Both inner and outer pins are placed below the wafer stripe and also above it to allow bending in both directions, upwards

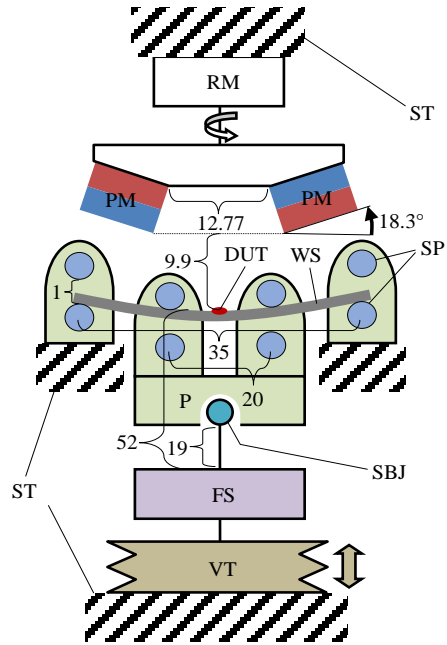

FIGURE 7: Measurement concept. RM (rotation motor), PM (permanent magnets), DUT (device under test) is on top side of WS (wafer stripe), SP (steel pins), P (moving piston), SBJ (spherical ball joint), FS (force sensor), VT (vertical translation stage), ST (stator). All lengths in mm. [Figure in black & white]

and downwards, for tensile and compressive stress on the top side of the silicon. The centers of the outer pins are $L_o$ = 35 mm apart, the centers of the inner pins are $L_i$ = 20 mm apart. Together with the large thickness of the silicon stripe (T = 0. 7397 mm) the resulting deformation is small up to considerable stress levels of $\sigma = \pm 160$ MPa. Beyond 160 MPa the wafer stripes tend to break, depending on the quality of their sawing edges. According to slim beam theory at small deflection the "stress per applied bending force" calibration factor is given by [31]

$$\frac{\sigma_{xy}}{F} = \frac{3}{4} \frac{L_o - L_i}{T^2 W} \frac{z}{T/2} \cong 2.05 \quad \text{MPa/N,} \tag{14a}$$

where $F$ is the total bending force and $z$ is the distance of the Vertical Hall device from the neutral fibre, $z = T/2 - 2$ μm. Equation (4.2) neglects the top layers. A finite element simulation with a 12.3 μm thick top layer gives a slightly smaller calibration factor.

$$\frac{\sigma_{xy}}{F} = 2.00 \quad \text{MPa/N} \tag{14b}$$

The total bending force is measured by a force sensor 8431-5050 from Burster Präzisionsmesstechnik (Gernsbach, Germany). The force sensor is mounted on a vertical translation stage HUMES 100-30-HiSM-IMS from OWIS (Staufen, Germany). For bending action the piston is moved up or down 0.5 mm by the vertical translation stage. The displacement of the piston is recorded. The spacing between upper and lower steel pins is 1 mm. Therefore no bending force acts on the wafer stripe during a piston stroke of 2*(1-0.752) =0.496 mm around the initial position. A slackness of the ball joint between force sensor and vertical translation stage could not be noticed, but it may well be hidden in a tolerance of the 1 mm gap between upper and lower steel pins.

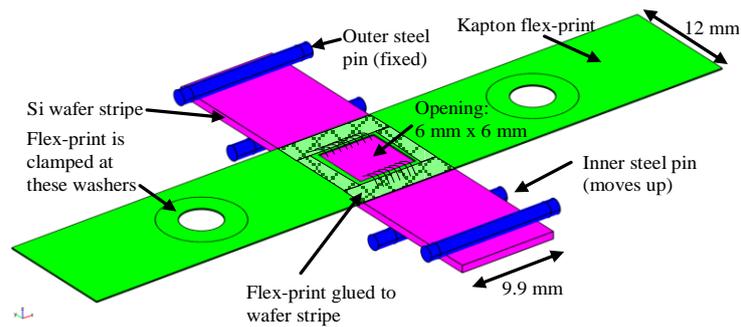

FIGURE 8: Schematic drawing of the wafer stripe with movable inner and fixed outer steel pins to apply the bending force. A flexible Kapton foil is glued to the wafer stripe. The current traces on the Kapton foil are not shown. Electrical contacts are made with bond wires.
[Figure in black & white]

Electrical connections are made via bond wires and a flexible Kapton foil with fine 35 µm thick conductor traces (see Figure 8). The foil is glued to the wafer stripe with a soft glue SMA10SL from Electrolube (Ashby de la Zouch, UK). The bond wires are covered with one drop of soft silicon gel to protect them. FEM simulations showed negligible influence of the foil and the glue on the stress pattern on the Hall devices. Each end of the foil is bolted to the stator of the bending fixture with a screw to decouple forces from cables on the Kapton foil. Thereby the foil provides enough slackness to avoid any tension during wafer bending.

A diametrical magnetic field acting on the Hall devices is generated by two NdFeB block magnets with edge lengths 10 mm x 40 mm x 5 mm, magnetized along the 5 mm edge. They are tilted around their 40 mm edge by 18.3° and spaced apart 12.77 mm in x-direction (see Figure 7). On the lower surface of one magnet there are magnetic north poles and on the lower surface of the other magnet there are magnetic south poles. This gives semi-circular magnetic flux lines which cross the rotation axis at 90°. The best homogeneity of the magnetic field is in the saddle point 3.9 mm below the magnets. However, at zero stress we placed the Hall devices 9.9 mm below the magnets due to space restrictions for the upper steel pins. At this location the horizontal flux density component is ~41 mT. Due to symmetry the vertical magnetic field component vanishes right on the rotation axis. However,

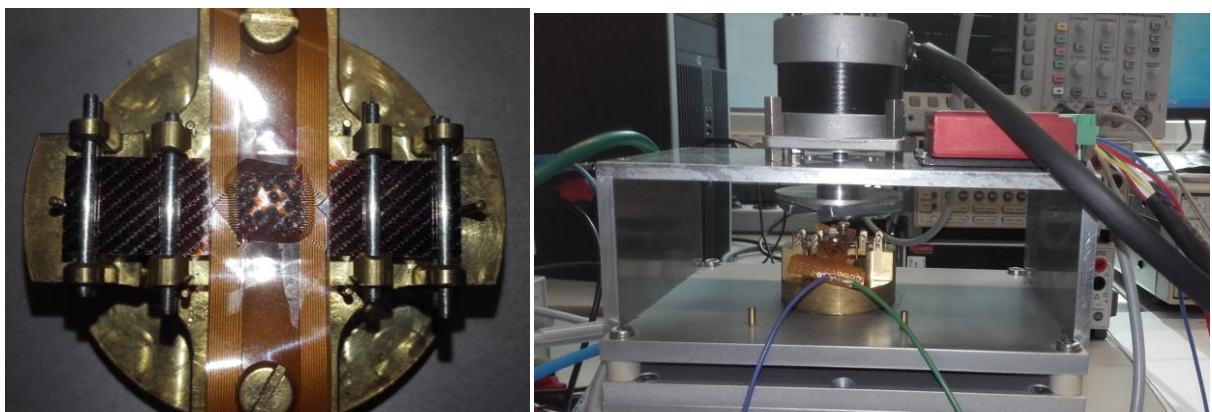

FIGURE 9: Photographs of the text fixture with a wafer stripe. The upper photo shows the wafer stripe with black polyimide coating and rows of test chips aligned at 45° to the stripe. The upper steel pins and the brass frames for the outer stator and the inner movable piston are visible. The Kapton foil has fine conductor traces for electrical contact of the center chip. Its bond wires are protected with a drop of glue. Note also the brass screws bolting the foil to the fixture to avoid forces from the cables onto the wafer stripe. The lower photo shows the two permanent magnets above the wafer stripe and the stepper motor for rotation. The vertical translation stage (not shown) is below the main brass block. [Figure in black & white]

the Hall devices are shifted in lateral direction by a few fractions of a millimeter according to the chip layout (cf. Figure 11) and the placement tolerances of the wafer stripe. 0.5 mm off the rotation axis the vertical field is less than 4.7 % of the lateral one. The maximum angular deviation of the magnetic flux lines is only 0.22° within a 4 mm diameter spherical volume around the point of nominal zero stress position. All these parameters of the magnetic field were found by a 3D magneto-static finite element simulation which assumes homogeneous magnetization with Brem = 1.17 T in the permanent magnets and a recoil permeability of 1.16.

The magnets are mounted to an aluminum holder, which is fixed to a rotatable shaft. The shaft is driven by a four-phase hybrid stepper motor with 1.8° step angle [32]. The rotational position is measured back by an angle encoder with a resolution of 8192 steps per revolution (13 bit). Unfortunately the absolute accuracy of the encoder was only +/-0.35° [33]. The rotation axis intersects the wafer stripe orthogonally at the location of the Vertical Hall devices – this holds for zero stress but also for arbitrary bending of the wafer stripe as long as the inner steel pins are perfectly centered between the outer steel pins and the forces applied to left and right steel pins are identical.

Figure 9 shows two photographs of the text fixture with a wafer stripe.

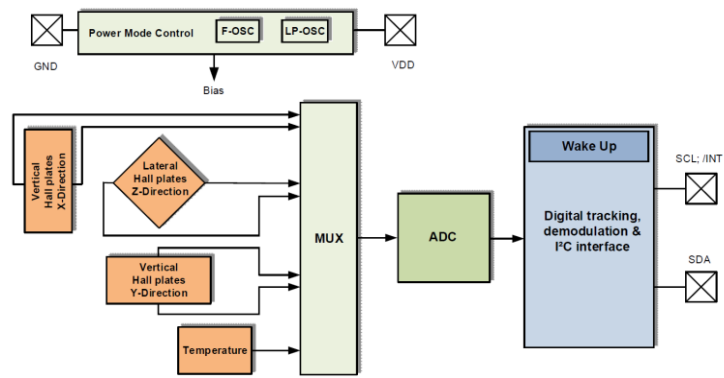

FIGURE 10: Concept diagram of the device under test. The sensor circuit has power supply, oscillator, and control circuits to operate the Hall devices and a temperature sensor in a time-multiplexed way. The signal path converts analog voltages into digital code. The temperature compensation is done in the digital domain, and all output signals are provided via a digital interface. [Figure in black & white]

### 4.2 Description of the device under test with Hall plates and Vertical Hall devices

Figure 10 shows the concept of the electronic circuit of the device under test. It is a commercially available 3D-Hall sensor with the layout given in Figure 11. Two groups of Vertical Hall devices are aligned along x- and y-directions (see Figure 12). Both are operated in a time multiplexed way with no power down or reset between time slots of x- and y-processing. They are supplied by the same bias current and their output signals are processed by the same preamp and analog-digital converter (ADC). Thus, any difference of readout values of x- and y-channels should come from the Vertical Hall devices. Therefore, it is possible to study the stress drift of the Hall signals despite the considerable complexity of the overall circuit. The Vertical Hall devices operate in a spinning current Hall probe scheme, which forces constant bias current and taps output voltage at alternately swapped pairs of contacts per device in order to reduce offset errors. The supply contacts of the Hall devices are switched at a frequency well above 10 kHz. Thus, transients of stress, temperature, and magnetic

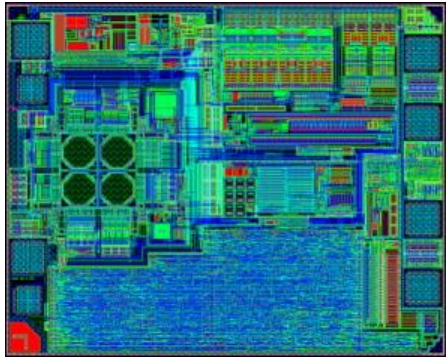

FIGURE 11: Layout of the device under test. A quadruple of Hall plates and the Vertical Hall devices are left of the chip center. The circuit supplies the Hall devices with adequate current decoupled from the chip supply voltage, amplifies their signals, removes their offset by a time discrete spinning current scheme, compensates their temperature drift, converts the signals to digital code, and outputs it through an I²C interface. [Figure in black & white]

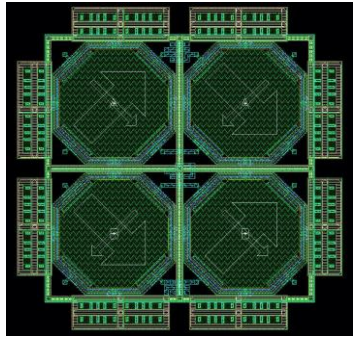

FIGURE 12: Zoom into the layout of the device under test (with top metal layer removed). A quadruple of Hall plates detects vertical magnetic field $B_z$. It is surrounded by 16 Vertical Hall devices (left and right columns) for detection of $B_x$ field and 16 Vertical Hall devices (top and bottom rows) for $B_y$ field. During the measurement we apply a magnetic field which is nearly parallel to the wafer surface with a very small $z$ component. [Figure in black & white]

field caused by the measurement equipment are significantly slower. The circuit of the test device also comprises an on-chip temperature sensor and the x-,y-, and z-channels are temperature compensated with an accuracy better than 0.1 %/°C. During operation and data transfer the test device operates at a supply voltage of 3.5 V (max) with a current drain of 5 mA (max). This may lead to moderate self-heating of the device under test (~1°C). The digital output signals have a quantization of 65 µT/LSB (LSB means least significant bit), but the measurement resolution was increased beyond that quantization limit by averaging 16 samples per data point.

### 4.3 The experimental procedure and recorded data

The following sequence of operations was executed: The piston was first moved up in small steps for tensile stress on the Hall devices, then it was moved down and up and finally down again. The up-movement towards the magnets produces tensile (= positive) stress on the top of the silicon stripe and the measured force is positive, too. At each stress value the magnet was rotated by 360° in 100 steps of 3.6° (which means exactly two steps of the stepper motor) and then the magnet was rotated back in its original 0° position again. The output signals of the Hall sensor circuit were recorded during the forward rotation only. A discrete Fourier transform (DFT) was applied to the data versus rotation angle, and the phases of the first harmonics were computed. The difference of phases of both Vertical Hall device signals should be 90°, but at mechanical stress it deviates by an amount, which is called orthogonality error.

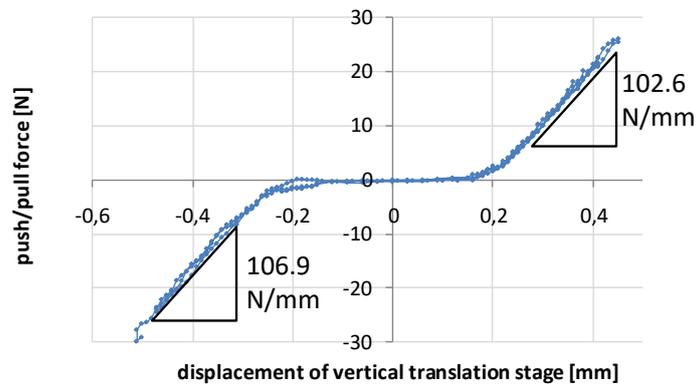

FIGURE 13: Sequence of measurements of wafer bending experiment with recorded bending force versus displacement of vertical translation stage. Positive displacement means that the piston was lifted up. This gives tensile stress in $[010]$ crystal direction on the Hall devices, and the force sensor outputs positive bending force. In the interval (-0.25 mm, 0.2 mm) the force vanishes, because the spacing of the upper and lower steel pins is larger than the wafer thickness. The point of zero displacement is arbitrary (because no provisions were made to detect whether the movable inner steel pins are at the same vertical position as the outer steel pins). The displayed force is the output signal of the force sensor minus its value at the start multiplied by the force transducer constant. The initial value comprises the offset error of the force sensor and the weight of the piston. [Figure in black & white]

Figure 13 shows a plot of the force measured by the force sensor versus displacement of the vertical translation stage. Each dot represents a full 360° rotation of the magnet. Near the origin, there is a displacement interval of length 0.466 mm, where the force is nearly constant: there we nulled the force sensor. This region of zero stress is consistent with the 1 mm gap between upper and lower steel pins and with the wafer thickness. Outside this 'dead zone' the slopes of the curve show a stiffness of 102.6 N/mm and 106.9 N/mm. An FEM simulation of the stiffness of the wafer stripe gave

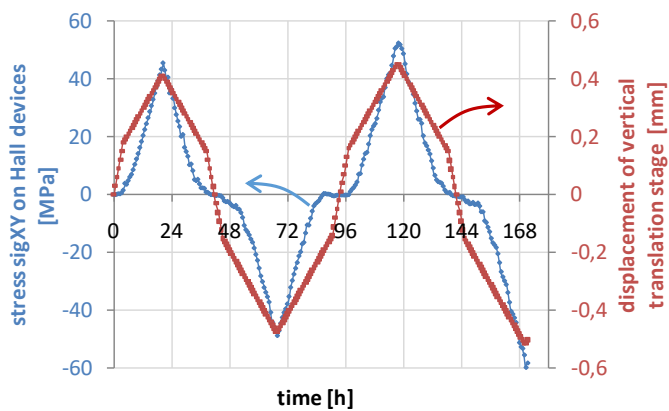

FIGURE 14: Same measurement procedure as in FIGURE 13, yet plotted versus time. The stress was obtained via the measured force and (4.3). [Figure in black & white]

a larger value of 130 N/mm. Thus it seems that the force sensors and other mechanical parts are also subject to small deformation, which varies between push and pull. The graph also shows a tiny hysteresis which may come from the friction of the piston in its bore or from the force sensor. Figure 14 shows the displacement of the vertical translation stage versus time. The measurement comprises 253 stress levels, where a full rotation was recorded during 41 minutes. This amounts to 172 hours, which is 7 days and 4 hours. In the zero stress region larger increments of displacements were used to speed up. The figure also shows the in-plane shear stress versus time with the stress derived from the measured force. Figure 15 shows the temperature of the device under test during this measurement as measured by the on-chip temperature sensor. During the first ramp up of stress, the temperature was ~23.7°C. For the rest of the measurement the mean temperature during the 360° rotations was from 21°C to 22°C. The temperature fluctuation was ~1°C during each 360° rotation of the magnet. If we discard the data of the first rising stress cycle, a fit gives -0.004 °C/MPa with a low

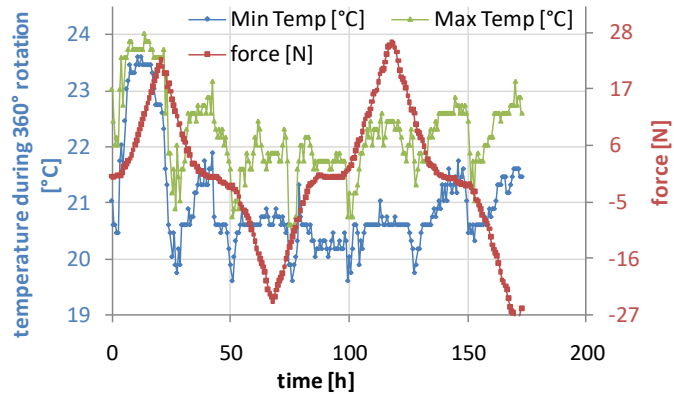

FIGURE 15: Temperature and bending force plotted versus time during the same measurement procedure as in Figures 13 and 14. The temperature was measured by the temperature sensor on the test chip on the wafer stripe. Min and Max denote the minimum and maximum temperature during the 360° rotation of the magnet. The temperature varies over the 41 minutes long 360° rotation but it also varies versus bending force, probably because of self heating and varying contact force between wafer stripe and steel pins. [Figure in black & white]

correlation factor of 0.035. This means that the measured stress dependence of the temperature sensor is mainly due to temperature fluctuations during the measurement, and the real stress dependence of the on-chip temperature sensor is very likely even lower.

We checked if the Kapton foil provides for enough slackness by removing the outer steel pins and moving the piston in the displacement coordinate from -0.53 mm to +0.53 mm. There the force sensor recorded a slope of 0.19 N/mm, which gives an error of less than 0.2% (cf. Figure 13). For displacements larger than +0.56 mm we noted a steep increase in measured force, because then the Kapton foil belts down the wafer stripe onto the piston. For displacements below -0.55 mm we note a soft decrease of force, because then a growing portion of the weight of the piston is lifted up by the Kapton foil.

In a similar experiment we also checked the homogeneity of the applied magnetic field. There we rotated the magnet until the x-signal had a maximum. Then the y-signal was close to zero. Displacement of the piston from -0.6 mm to +0.6 mm led to a 0.1° increase of the in-plane angle (*i.e.* the angle detected by x- and y-Vertical Hall devices). Thereby the magnitude of the magnetic field increased by 7%, which proves that the stress free position is not in the saddle point of the magnetic field.

At each stress level the data of the Hall devices were recorded. Figure 16 shows these exemplary raw signals for three levels of in-plane shear stress $\sigma_{xy}$ (-45 MPa, 0 MPa, and +45 MPa). The amplitudes of the X- and Y-signals match up to 0.4% but the Z-signal is 27 times smaller. The phases of the X- and Y-signals are shifted into opposite directions by the stress, while the phase of the Z-signal remains constant versus stress. This is in accordance with our theory.

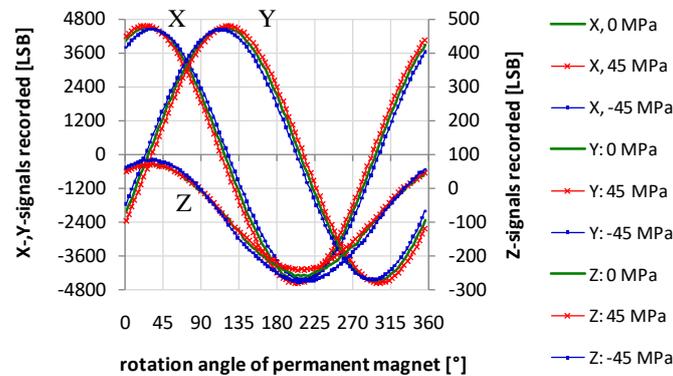

FIGURE 16: X-, Y-, Z- signals recorded during 360° rotations of the magnet at three different levels of shear stress 0 MPa, 45 MPa, -45 MPa. Stress moves the X- and Y-signals in opposite directions. At zero stress the Z-signals have accidentally the same phase as the X-signals. [Figure in black & white]

An extraction of the phases of the Hall devices versus all applied stress levels gives the plot in Figure 17. The phase of the X-Vertical Hall devices rises with 99.6 °/GPa whereas the phase of the Y-Vertical Hall devices decreases with -97.8 °/GPa. The phase of the Z-Hall plates remains constant versus stress, but it jumps by ~1° when we switch from compressive to tensile. The reason for this jump is probably a misalignment of upper and lower inner steel pins by 1° due to inaccuracies of the manufacturing process.

Subtraction of phases of both Vertical Hall devices gives the orthogonality error with a least square fit result of 0.1974°/MPa (Figure 18). According to FEM simulations we expect 0.233°/MPa. Hence, the measured effect is strong, but it is 15% smaller than expected. The discrepancy seems to come

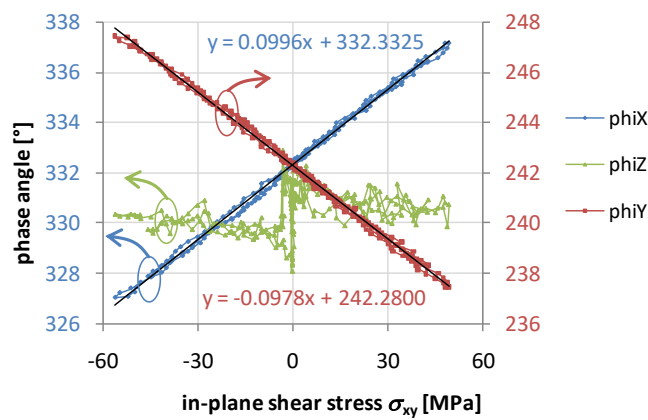

FIGURE 17: Phase angles of the sinusoidal signals of x/y-Vertical Hall devices and z-Hall plates during rotations of the magnet at varying shear stress. PhiX increases and phiY decreases significantly versus stress, while phiZ remains constant. The small 1° jump in phiZ between compression and tension seems to come from upper and lower inner steel pins being not perfectly parallel. The permanent magnet was accidentally aligned at 332.3° to the x-axis of the sensor chip. The orthogonality error at zero stress is small: 0.05°. [Figure in black & white]

from inaccurate modeling of doping profiles of nCMOS-wells through small LOCOS openings in our BiCMOS technology. In Figure 18 the sum of phases of both Vertical Hall devices seems to be constant versus stress, as predicted by theory. The small drift of this sum over the entire measurement (~0.3°) may be caused by the measurement procedure. Perhaps the Kapton foil slightly pulled the wafer stripe during the zero-stress transitions. Or the misalignment of lower and upper inner steel pins and some slackness of the ball joint gave rise to minute changes of wafer stripe position over several stress cycles. Whenever stress starts at zero and builds up to its tensile or compressive peak value, the sum of phases remains fairly constant versus stress – only at the subsequent relaxation of stress we note a tiny drift around 0.1°. Anyway, the drift in the sum of phases is smaller than 1.5% of the drift in orthogonality. In contrast to [19] we saw only tiny orthogonality errors at zero stress.

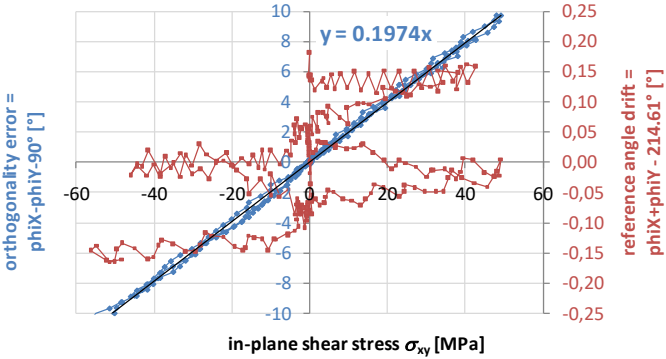

FIGURE 18: Orthogonality error of x-y-Vertical Hall devices amounts to 0.1974°/MPa of sigXY. The sum of phase angles phiX+phiY drifts from 0.15° to -0.15° during the measurement, but this seems to be only measurement artifacts. [Figure in black & white]

### 4.4 A second experiment on naked Vertical Hall devices

We did a similar measurement on naked Vertical Hall devices, where supply current was provided by a Keithley sourcemeter and switched to the appropriate contacts via a relais matrix while the output voltage was sampled by a multimeter (K2001 from Keithley Instruments Inc., Cleveland, OH, operated in 100 mV range). A slow spinning cycle with ~1 s per spinning phase was executed and the voltages were summed up numerically in a post-processing step. Two types of Vertical Hall devices were studied – both are in another BiCMOS technology node (with smaller feature size) than the one in the preceding section, yet all technologies had ~5 µm deep n-epitaxial Hall regions isolated in lateral direction by deep trenches and in vertical direction by an n-buried layer in a (100) silicon p-substrate. The layouts are given in Figure 19. The devices are of similar width but their length varies

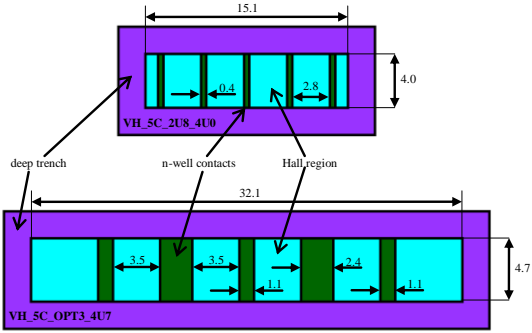

FIGURE 19: Layouts of two Vertical Hall devices, which were measured without additional on-chip circuit in the second experiment. All lengths are in µm. [Figure in black & white]

by a factor of two. The mechanical test fixture was also modified: it used 1.6 mm diameter nonmagnetic tungsten pins instead of grade A4 steel pins at a spacing of 15 mm (inner ones) and 30 mm (outer ones). The spherical ball joint was replaced by a nonmagnetic part. The wafer stripes were smaller with larger aspect ratio: 32 mm x 4.95 mm. Their thickness was again 752 µm. Figure 20 shows the measurement result of the orthogonality drift versus shear stress. It is 0.1880°/MPa and 0.2207°/MPa, respectively. Both values are similar to the 0.1856°/MPa in Figure 18.

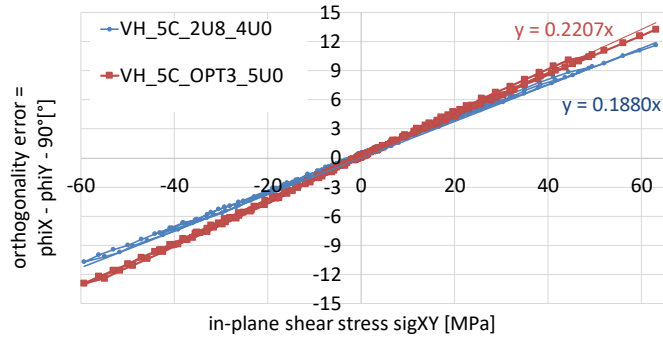

FIGURE 20: Orthogonality between x- and y-Vertical Hall devices from Figure 19. [Figure in black & white]

## 5 Discussion

### 5.1 Accuracy of measurements

The alignment accuracy of the wafer stripes w.r.t. the crystal axes is mainly given by the alignment of the wafer flat in commercial (100)-silicon wafers. It was specified by 1°, but usually it is better than 0.5°. The dicing accuracy is better than 5 µm over a 200 mm diameter wafer, which means < 0.0015°.

The bending force is applied to the wafer stripe via steel pins. The steel is grade A4, which is slightly ferromagnetic. Originally, we intended to use non-magnetic brass pins, but finite element simulations showed that steel is required for bending forces up to 40 N. At this maximum force the von Mises stress in the contact area of the inner steel pins reaches 75 MPa, which is near the fatigue strength of cyclic loading. Inevitably, the steel pins are slightly deformed due to the bending force (the upper pins are more deformed because the lower pins are supported along their full length). This exerts more contact force onto the wafer stripe near its long edges than near its center line, which leads to inhomogeneous mechanical stress on the wafer stripe near the steel pins. The stress is homogeneous in the center between the steel pins if they are sufficiently distal. Within a circle of 4 mm diameter around the center of the stripe and at a depth of 2 µm below the top surface of the silicon $\sigma_{xy}$ is homogeneous up to -1%/+1.8% and the other stress components are $(\sigma_{xx} + \sigma_{yy})/\sigma_{xy0} = 2.222 \ldots 2.254$, $(\sigma_{xx} - \sigma_{yy})/\sigma_{xy0} = -2.8\% \ldots 0\%$, , with $\sigma_{xy0}$ being the shear stress in the center.

Note that the 9.9 mm width of the stripe is not negligible against the 20 mm spacing of the inner steel pins, and this generates normal stress not only along the length but also along the width direction: transverse stress is 5.5% of longitudinal stress according to FEM simulation. This reduces the calibration factor (14b) by another 5.5% which we accounted for.

We also assessed the effect of the layers on top of the silicon wafer stripe on the mechanical stress per applied bending force. The mechanical parameters of silicon were taken from equation (8) in [34]. The stack of thin layers was modeled as a single layer with Young's modulus E = 50 GPa and Poisson ratio equal to 0.334. These effective parameters were obtained by the following rule of mixture: summing up the parameters of the individual layers weighted with the volume fraction of each layer. The parameters of the individual layers are difficult to estimate, because they depend on layer thickness, crystalline structure (single vs. poly-crystalline vs. amorphous), manufacturing process (LPCVD vs. PECVD with different deposition temperatures), anneal temperatures, hydrogenization etc... Therefore, the accuracy of the 50 GPa is only modest. Yet, the top layer was 60 times thinner and three times softer than the bulk silicon and consequently its exact material parameters are of little relevance for the mechanical stress calculation in the silicon. We checked this with an analytical calculation of the stress in a laminate composed of two layers according to [27] (see Appendix C): if the effective top layer had a Young's modulus of 20 GPa or 80 GPa instead of the nominal 50 GPa assumed in (14b) the stress on the Vertical Hall devices would change by 1.6% for the same bending force. If we ignore the top layers altogether, *i.e.*, we set their Young's modulus zero, we would overestimate the stress by 2.7%. Conversely, if we assume the top layers to be as stiff as bulk silicon, we would underestimate the stress by 4.1%. Interestingly, this effect of top layers on the stress in wafer bending experiments seems not to have received adequate attention in the literature so far.

The tubs of the Vertical Hall devices are made of n-epitaxial layer, which extends from the top of the single crystal silicon into a depth of ~5 µm. Doping and current flow density are inhomogeneous versus depth coordinate. Bending stress is zero in the neutral fibre and builds up linearly towards the top surface. Therefore it changes by 1/(739.7/2) = 0.27% per micro-meter. We consider the mechanical stress in a depth of 2 µm below the top of the bulk silicon as being relevant for the Vertical Hall devices. This reduces the "stress per applied bending force" calibration factor by 0.54%. Due to the poorly defined top layers and the finite thickness of the Vertical Hall devices we prefer to use wafers with maximum thickness, which is ~0.75 mm.

Once the wafer stripe is placed into the bending fixture its clearance is only +/-0.05 mm in width direction and +/-0.7 mm in length direction. The mono-reticle has only one sort of chip. The center chip is connected to conductor traces on the Kapton foil. The Hall devices are placed in the center of the wafer stripe with a placement accuracy of better than 1 mm.

The maximum allowed force on the force sensor is 50 N (compressive and tensile). The weight of the piston is ~1.2 N. The zero-point error of the force sensor is irrelevant, because it is nulled at the start of each measurement cycle. The force sensor has a voltage output and its transducer gain is 251.5 µV/N at a supply voltage of 5 V, which we calibrated with a one liter water bottle with an estimated accuracy of +/-4%. The linearity is better than +/-0.075 N. The hysteresis is less than 0.1 N. The temperature stability of the output is 0.015 N/°C from -55 °C to +120 °C. The output voltage of the force sensor was measured with a K2000 voltmeter in 100 mV range (from Keithley Instruments Inc., Cleveland, OH). The force sensor is cylindrical with 19 mm diameter and 13 mm height. It is ferromagnetic, but due to its rotational symmetry and distance of 52 mm to the magnet it should not alter the direction of the magnetic field applied to the wafer stripe during bending action.

The vertical translation stage can carry a maximum weight of 120 N and it provides a displacement sensor with an accuracy of 5 µm (bidirectional).

The inner steel pins are mounted on a cylindrical brass piston which rests in a bore with 0.2 mm clearance. The clearance of the piston and the 0.1 mm clearance of the wafer stripe between the inner steel pins of the bending fixture guarantee that the steel pins are perpendicular to the length axis of the stripe with a deviation of less than 1°. This small misalignment corresponds to an error of -0.03% in shear stress on the Hall devices in the center of the beam. If the beam rotation doubles from 1° to 2° also the error in shear stress doubles to -0.066% according to FEM simulation. Due to the clearance of the piston the inner steel pins may also be laterally misaligned against the outer steel pins by 0.2 mm. This eccentricity overestimates the shear stress by 1% according to FEM simulation. Summing up all these errors gives only a moderate absolute accuracy of +/-10% for the shear stress.

The movable piston with inner steel pins is coupled to the force sensor by a rod end with ball joint, which prevents different forces on left and right steel pins. The ball and the rod end are strongly ferromagnetic. Its lower part is a cylinder (19 mm long, 9mm diameter) with inner thread. With its rotational symmetry, it should not affect the rotation angle of the applied field. The same applies for the ball in the joint. The upper part of the rod end is a cylindrical ring, which holds the steel ball. Its cylinder axis is parallel to the wafer stripe. This cylindrical ring has inner and outer diameters of 13 and 17 mm, respectively, and a thickness of 6 mm. The ball of the joint has 13 mm diameter with a 5 mm diameter hole through which a brass pin bolts it to the piston.

Due to the curvature of the bent wafer stripe, the Hall devices move slightly more than the piston: According to FEM simulation a force of 40 N moves the piston 306 µm and the Hall devices 468.3 µm. Hence, a total recorded displacement of 0.883 mm means that the Hall devices moved 0.496 mm without bending and (0.883 – 0.496)*468.3/306 = 0.592 mm due to bending action. This gives a total stroke of 1.088 mm by which the Hall devices move relative to the applied magnetic field.

The only relevant ferromagnetic parts without rotational symmetry with respect to the rotation angle of the applied magnetic field are the steel pins, the hole in the ball of the ball joint and its cylindrical ring of the rod end. A finite element (FEM) simulation shows a deviation of the magnetic field angle on the Hall devices of 0.1° at rotational positions 0° and 180° and -0.1° at rotational positions 90° and 270°. Thus, the deviation is at the second harmonic so that it is irrelevant if we evaluate the phase of the first harmonic. Moreover, this deviation increases only by 0.05° if the DUT approaches the ball joint by 1 mm.

Unfortunately, the rotational movement of the magnet was recorded with poor accuracy of the encoder (see above: +/-0.35°). Indeed, if we take the naked Hall signals, subtract their offsets, divide by the amplitudes of their first harmonics, correct for the phase angles, and compute the arctangent of both, the result differs from the encoder angle by +/-0.3°. The largest noise spikes on this residual curve are below 0.1° so that the main portion seems to come from the limited accuracy of the encoder. However, we were not interested in the higher harmonics of the angle error but in the phase of the fundamental frequency of the sine/cosine Hall signals. Therefore, errors of the encoder in a local vicinity around any rotational position should be irrelevant. Conversely, errors of the encoder with a 360° periodicity would show up in the extracted phases of the Hall signals, yet they should not change versus applied mechanical stress. Therefore, the drift of the phases should not suffer unduly from poor encoder accuracy.

We chose angular steps of 3.6°, because this is twice the minimum step size of the stepper motor. However, due to manufacturing tolerances of the motor not all steps are exactly equal. The encoder measures a standard deviation of 0.05° for the step size. We computed the phase angles with discrete Fourier transform (DFT) implemented in EXCEL. Ordinary DFT implicitly assumes that data is sampled on a regular grid of angles. Thus the spread of angle steps gives a small error in the DFT. The phases of a few rotations were also computed with MATHEMATICA, which offers a fit procedure that can handle irregularly sampled data. The differences in the phases between EXCEL and MATHEMATICA was less than 0.025°, the orthogonality differed by no more than twice that value, and the sum of both phases differed by 0.002°. The largest values of orthogonality plotted in Figure 18 are around 10° and therefore the 0.05° error caused by irregular sampling grid is only 0.5% of the observed orthogonality drift for a single large stress value. Since the least square error fit line in Figure 18 uses 250 stress levels this type of error should be negligible in the resulting slope of 0.1974°/MPa.

**5.2 Practical relevance of shear stress related crosstalk**

With ~160%/GPa the crosstalk between x- and y-components of the magnetic field is very sensitive to shear stress for Vertical Hall devices. For Hall plates the crosstalk between in-plane and out-of-plane magnetic field components is ~25 times less sensitive to shear stress. Thus, the apparent angular alignment of Vertical Hall devices is stable up to ~0.1°/MPa and ~0.004°/MPa for Hall plates. In both cases the specific combination of piezo-Hall coefficients for (100) wafer orientation is responsible for the magnitude of the effect. Moreover, large stress levels well above 100 MPa may occur on the top surface of silicon chips in plastic encapsulated packages. Global stress patterns originate from mismatch in thermal coefficient of expansion of leadframe, silicon, and mold compound, but also from moisture swelling and chemical cure shrink of mold compound and die attach adhesives. Local inhomogeneous stress may come from deep trench isolation, interconnect layers, and bumps for flip chip assembly. On the other hand, the relevant stress components for crosstalk are in-plane shear stress $\sigma_{xy}$ for Vertical Hall devices and out-of-plane shear stresses $\sigma_{xz}, \sigma_{yz}$ for Hall plates. Luckily, the laminar structure of common packages generates only in-plane normal stress $\sigma_{xx}, \sigma_{yy}$ on the major portion of the chip surface. Significant shear stress occurs only on the circumference of the chip and particularly in its corners. However, Hall sensor chips are usually small (~2 mm²) and the circumferential region is roughly as wide as the chip thickness (principle of Saint Venant). Therefore it is good practice to place the Hall devices in the center of the die, to use symmetrical leadframes and packages, and to avoid thick dies. Observations from stress test chips and commercial Hall sensors suggest that in spite of these precautions shear stresses of a few mega-Pascals on the Hall devices are unavoidable in standard packages. Therefore, prior art magnetic angle sensors with Vertical Hall devices will exhibit stress related angle errors of a few fractions of a degree via two mechanisms: (i) the phase shift between signals of orthogonal Vertical Hall devices deviates from 90°, and (ii) the phase of any of the two Vertical Hall device signals is not constant versus stress at a 0° reference angle. We call the former orthogonality error and the latter drift of reference angle.

For continuous rotation, a smart angle sensor can perform an auto-calibration algorithm during operation. There it observes the signals of both Vertical Hall devices over a full revolution of 360° and extracts the orthogonality error from these data. From now onwards it can take account for the orthogonality error and cancel it out. But how about the drift of reference angle? When exposed to shear stress the sensitive directions of orthogonal Vertical Hall devices converge or diverge depending on the sign of the stress. It means that one of them is rotated in clockwise direction and the other one in counter-clockwise direction. Due to this strong correlation of both Vertical Hall devices the direction $\vec{n}_x + \vec{n}_y$ in the center between both device orientations remains constant versus stress. A smart angle sensor uses this stable direction $\vec{n}_x + \vec{n}_y$ as a reference – in other words, it accounts for the drift of the reference angle, which is half of the orthogonality error.

For applications without continuous rotation, such as windscreen wipers or throttle valve position sensors, or at start up (before a full revolution is completed), there is a limit to auto-calibration routines. In such cases we propose a stress compensation circuit similar to [35]: a smart angle sensor can measure the shear stress on its Vertical Hall devices. It estimates the change in sensitive direction of the Hall devices and accounts for it *e.g.* in the digital part of the signal conditioning circuit.

With auto-calibration or stress compensation methods it should be possible to provide rugged, tiny, and cheap magnetic angle sensors with temperature and lifetime drifts below 0.1°.

**6 Conclusion**

We showed in this paper that shear stress can lead to a considerable crosstalk of orthogonal magnetic field components in (100)-silicon Hall effect devices. Up to our knowledge, this effect has not been mentioned before. It is negligible for conventional Hall plates, but it needs to be considered for premium Vertical Hall devices in automotive and industrial applications. To get a qualitative understanding, we applied the classical piezo-resistance and piezo-Hall theory and performed 3D finite element simulations. For an experimental proof we used a simple four-point bending fixture for wafer stripes and exposed it to a rotating magnetic field. Measurements were done on naked Vertical Hall devices and on a commercial smart 3D-Hall sensor comprising elaborate biasing and signal conditioning circuits (including analog-to-digital conversion of the output signals). All results match reasonably well. They consistently show a crosstalk of ~0.16%/MPa of in-plane shear stress. In a rotating in-plane magnetic field this shifts the phase of the signal of the Vertical Hall device by 0.092°/MPa, and in a magnetic angle sensor it leads to an orthogonality error of 0.18°/MPa. This crosstalk can be reduced by one order in magnitude with a simple cost efficient stress compensation circuit, which continuously measures the shear stress and accounts for it in the signal conditioning part of a smart Hall sensor. With these provisions, we are confident that future magnetic angle sensors will have a lifetime stability better than 0.1° in spite of their miniature low cost plastic packages.

**Conflicts of interest:**

The authors declare no conflicts of interest regarding the publication of this paper.

**References**

[1] Schott, C., & Popovic, R. S. (1997, June). Linearizing integrated Hall devices. In *Proceedings of International Solid State Sensors and Actuators Conference (Transducers' 97)* (Vol. 1, pp. 393-396). IEEE. **DOI:** 10.1109/SENSOR.1997.613667

[2] Hendrik Husstedt, private communication.

[3] Munter, P. J. A. (1990). A low-offset spinning-current Hall plate. *Sensors and Actuators A: Physical*, *22*(1-3), 743-746. **DOI:** 10.1016/0924-4247(89)80069-X

[4] Husstedt, H., Ausserlechner, U., & Kaltenbacher, M. (2011). In-situ analysis of deformation and mechanical stress of packaged silicon dies with an array of Hall plates. *IEEE Sensors Journal*, *11*(11), 2993-3000. **DOI:** 10.1109/JSEN.2011.2154326

[5] Popović, R. S. (1989). Hall-effect devices. *Sensors and Actuators*, *17*(1-2), 39-53. **DOI:** 10.1016/0250-6874(89)80063-0

[6] Ausserlechner, U. (2016). Hall effect devices with three terminals: Their magnetic sensitivity and offset cancellation scheme. *Journal of Sensors*, *2016*. **DOI:** 10.1155/2016/5625607

[7] Schott, C., Besse, P. A., & Popovic, R. S. (2000). Planar Hall effect in the vertical Hall sensor. *Sensors and Actuators A: Physical*, *85*(1-3), 111-115. **DOI:** 10.1016/S0924-4247(00)00328-9

[8] H. J. Carlin and A. B. Giordano, *Network Theory: An Introduction to Reciprocal and Nonreciprocal Circuits,* Prentice-Hall, New Jersey, NJ, USA, 1964.

[9] Ausserlechner U., (2004, October). Limits of offset cancellation by the principle of spinning current Hall probe. In *SENSORS, 2004 IEEE* (pp. 1117-1120). IEEE. **DOI:** 10.1109/ICSENS.2004.1426372

[10] Popovic, D. R., Dimitrijevic, S., Blagojevic, M., Kejik, P., Schurig, E., & Popovic, R. S. (2006, April). Three-axis teslameter with integrated hall probe free from the planar hall effect. In *2006 IEEE Instrumentation and Measurement Technology Conference Proceedings* (pp. 1812-1815). IEEE. **DOI:** 10.1109/IMTC.2006.328256

[11] Burger, F., Besse, P. A., & Popovic, R. S. (2001). Influence of silicon anisotropy on the sensitivity of Hall devices and on the accuracy of magnetic angular sensors. *Sensors and Actuators A: Physical*, *92*(1-3), 175-181. **DOI:** 10.1016/S0924-4247(01)00560-X

[12] Hälg, B. (1988). Piezo-Hall coefficients of n-type silicon. *Journal of applied physics*, *64*(1), 276-282. **DOI:** 10.1063/1.341422

[13] Ausserlechner, U. (2018). An Analytical Theory of Piezoresistive Effects in Hall Plates with Large Contacts. *Advances in Condensed Matter Physics*, *2018*. **DOI:** 10.1155/2018/7812743

[13b] Sunier, R., Taschini, S., Brand, O., Vancura, T., & Baltes, H. (2003, June). Quasi-analytical study of offset voltage due to piezoresistive effect in vertical Hall devices by mapping techniques. In *TRANSDUCERS'03. 12th International Conference on Solid-State Sensors, Actuators and Microsystems. Digest of Technical Papers (Cat. No. 03TH8664)* (Vol. 2, pp. 1582-1585). IEEE. **DOI:** 10.1109/SENSOR.2003.1217082

[14] Pfann, W. G., & Thurston, R. N. (1961). Semiconducting stress transducers utilizing the transverse and shear piezoresistance effects. *Journal of Applied Physics*, *32*(10), 2008-2019. **DOI:** 10.1063/1.1728280

[15] Kanda, Y., & Yasukawa, A. (1981). Hall-effect devices as strain and pressure sensors. *Sensors and Actuators*, *2*, 283-296. **DOI:** 10.1016/0250-6874(81)80048-0

[16] M. Baumann, Robust Piezoresistive CMOS Sensor Microsystem, thesis at IMTEK, Univ. of Freiburg, (2013)

[17] Ausserlechner, U. (2013). A theory of magnetic angle sensors with hall plates and without fluxguides. *Progress In Electromagnetics Research*, *49*, 77-106. **DOI:** 10.2528/PIERB13011011

[18] Demierre, M. (2003). *Improvements of CMOS Hall microsystems and application for absolute angular position measurements* (Doctoral dissertation, EPFL no. 2844, Lausanne, Switzerland). Section 6.1.3

[19] Beran, P., Stahl-Offergeld, M., Peters, V., Krause, D., & Hohe, H. P. (2018). Impact of contact misalignment on magnetic cross sensitivity of integrated vertical Hall sensors. *IEEE Transactions on Magnetics*, *55*(1), 1-4. **DOI:** 10.1109/TMAG.2018.2873238

[20] R. E. Newnham. "Properties of Materials. Anisotropy, Symmetry, Structure." Oxford Univ. Press, New York, 2005, chapter 10

[21] J. F. Nye. "Physical Properties of Crystals. Their Representation by Tensors and Matrices." Oxford Univ. Press, New York, 1985 chapter 8.2

[22] Keyes, R. W. (1960). The effects of elastic deformation on the electrical conductivity of semiconductors. In *Solid State Physics* (Vol. 11, pp. 149-221). Academic Press. **DOI:** 10.1016/S0081-1947(08)60168-X

[23] Price, P. J. (1957). The linear Hall effect. *IBM Journal of Research and Development*, *1*(3), 239-248. DOI: 10.1147/rd.13.0239

[24] Seeger, K. (1988). Semiconductor Physics, 4th edn., Springer Ser. Solid-State Sci.

[25] Onsager, L. (1931). Reciprocal relations in irreversible processes. II. *Physical review*, *38*(12), 2265. **DOI:** 10.1103/PhysRev.38.2265

[26] Mian, Ahsan, Jeffrey C. Suhling, and Richard C. Jaeger. "The van der Pauw stress sensor." *IEEE Sensors Journal* 6.2 (2006): 340-356. **DOI:** 10.1109/JSEN.2006.870140

[27] Thermal stress and strain in microelectronics packaging, J. H. Lau (ed.), van Nostrand Reinhold, NY (1993), chapter 2.

[28] Kanda, Y., & Suzuki, K. (1995, January). The piezo-Hall effect in n-silicon. In *22nd International Conference on the Physics of Semiconductors* (Vol. 1, pp. 89-92).

[29] Ausserlechner, U. (2018). An analytical theory of Hall-effect devices with three contacts. *Open Physics Journal*, *4*(1). **DOI:** 10.2174/1874843001804010014

[30] Ausserlechner, U. (2018). An Analytical Theory of the Signal-to-Noise Ratio of Hall Plates with Four Contacts and a Single Mirror Symmetry. *Journal of Applied Mathematics and Physics*, *6*(10), 2032-2066. **DOI:** 10.4236/jamp.2018.610174

[31] Beaty, R. E., Jaeger, R. C., Suhling, J. C., Johnson, R. W., & Butler, R. D. (1992). Evaluation of piezoresistive coefficient variation in silicon stress sensors using a four-point bending test fixture. *IEEE Transactions on Components Hybrids and Manufacturing Technology*, *15*(5), 904-914. **DOI:** 10.1109/33.180057

[32] motor no. 440-458, https://docs-emea.rs-online.com/webdocs/1579/0900766b81579a94.pdf

[33] no. 7951239 https://docs-emea.rs-online.com/webdocs/1583/0900766b81583409.pdf

[34] Hopcroft, M. A., Nix, W. D., & Kenny, T. W. (2010). What is the Young's Modulus of Silicon?. *Journal of microelectromechanical systems*, 19(2), 229-238. **DOI:** 10.1109/JMEMS.2009.2039697

[35] Motz, M., & Ausserlechner, U. (2017). Electrical compensation of mechanical stress drift in precision analog circuits. In *Wideband Continuous-time ΣΔ ADCs, Automotive Electronics, and Power Management* (pp. 297-326). Springer, Cham. **DOI:** 10.1007/978-3-319-41670-0_16

[36] Adel S. Sedra and Kenneth C. Smith, Microelectronic Circuits, Oxford Univ. Press 1998, 4[th] ed., ch. 3.3.2, eq. (3.20)

**Appendix A: Tilt of Hall plate due to junction-field effect according to Hendrik Husstedt**

The Hall plate has a metallurgical thickness $t_H = 900$ nm. It is n-doped with donor concentration $N_D$ and it is junction isolated, *i.e.* its surroundings is p-doped substrate with acceptor concentration $N_A$. When the pn-junction between Hall plate and substrate is reverse biased a depletion layer establishes. It reaches a depth $t_n$ into the Hall plate and a depth $t_p$ into the substrate. The depletion layer acts as isolation, thus it reduces the effective thickness of the Hall tub down to a value of $t_H - t_n$. [36] gives the value $t_n = \sqrt{2\varepsilon_{Si}(V_0 + V(x))/(qN_D)}/\sqrt{1 + N_D/N_A}$ with $\varepsilon_{Si} = 11.7 \times 1.04 \times 10^{-10} \text{Fm}^{-1}$. $V_0 = V_T \ln(N_A N_D / n_i^2) \approx 0.716$ V is the built-in voltage with the thermal voltage $V_T = kT/q$ (~26 mV at room temperature) and the intrinsic carrier density $n_i \approx 1.5 \times 10^{-21} \text{m}^{-3}$ at room temperature in silicon. *V(x)* is the reverse bias voltage across the junction as a function of the lateral position *x*. For $N_A = N_D = 2 \times 10^{21} \text{m}^{-3}$ we get $t_n \approx 0.18 \mu\text{m} \times \sqrt{0.716 \text{ V} + V(x)}$. *x* is the coordinate between two opposite contacts. It ranges from 0 to 75.5 μm. Since the contacts lead to a certain degree of current crowding, the voltage does not drop linearly along *x* – even if there is no junction field effect. However, Hall plates with good signal to noise ratio have fairly large contacts so that we may neglect current crowding. At the ground contact the depletion layer is 152 nm wide, at the positive supply contact it is 297 nm wide for a supply voltage of 2 V. Thus, the effective thickness of the Hall tub varies from 83% to 67% of the metallurgical thickness. A linear fit gives a tilt of $\tau = 0.109°$.

We checked this with a finite element simulation. The Hall plate had regular octagonal shape with edge length 36.4 μm. The contacts are rectangular 2.1 μm x 30 μm. The spacing of opposite contacts is 75.5 μm. The x- and y-axes run midway between the contacts. The thickness and the depletion layer is identical to above, whereby the bottom of the tub is at z = 0 and the top is at z = 0.9 μm. The mesh is critical for this type of problem: first we meshed the top surface with 267025 triangular elements, then we swept the mesh along the thickness direction in such a way as to obtain 48 layers with a special distribution of thicknesses near the bottom, where the depletion layer occurs (see Figure A1). Thus, the total number of elements was about 12.82 million and the degrees of freedom for the static conduction problem were 13.25 million. The conductivity tensor was modeled like this

$$\vec{\kappa} = \left(0.01 + \frac{0.99 \times z^{14}}{z^{14} + t_n^{14}}\right) \times \rho^{-1} \begin{pmatrix} 1 & \rho^{-1} C_H B_z & -\rho^{-1} C_H B_y \\ -\rho^{-1} C_H B_z & 1 & \rho^{-1} C_H B_x \\ \rho^{-1} C_H B_y & -\rho^{-1} C_H B_x & 1 \end{pmatrix} \quad \text{(A1)}$$

The pre-factor models the depletion layer by a smooth transition from 1% of the bulk conductivity at $z = 0$ via 50.5% at $z = t_n$ to greater than 99.7% at $z > 1.5 \times t_n$. 0 V and 2 V were applied to two opposite contacts, while the other contacts were floating. A sequence of calculations was done with the following magnetic flux densities: first $\vec{B} = (0.1\,\text{T}) \times \vec{n}_z$, then $\vec{B} = (-0.1\,\text{T}) \times \vec{n}_z$, then $\vec{B} = (0.1/\sqrt{2}\,\text{T}) \times (-\vec{n}_x + \vec{n}_y)$, and finally $\vec{B} = (0.1/\sqrt{2}\,\text{T}) \times (\vec{n}_x - \vec{n}_y)$. The first two cases have a magnetic field perpendicular to the Hall plate, the last two cases have a magnetic field along the direction of both supply contacts. Subtracting the voltages of the output contacts for positive and negative magnetic fields and dividing it by 0.2 T and 2 V gives the voltage related magnetic sensitivities of 63.71 mV/V/T for z-field and 84.5 µV/V/T for lateral field. This corresponds to a tilt of $\tau/2 = (84.5/63710) \times 180°/\pi = 0.076°$ of the sensitive direction of the Hall plate. This matches roughly the crude estimation given above: $\tau = 0.152°$ versus $\tau = 0.109°$. The current density averaged over the entire Hall tub gives a vector which is tilted 0.055° out of the top surface of the plate. This is very close to half of the tilt of the junction isolation interface. It was also checked that a magnetic field in the direction $\vec{B} = \pm(0.1/\sqrt{2}\,\text{T}) \times (\vec{n}_x + \vec{n}_y)$ does not generate any output signal. Thus, these results of a numerical simulation are in reasonable agreement with experimental evidence [2].

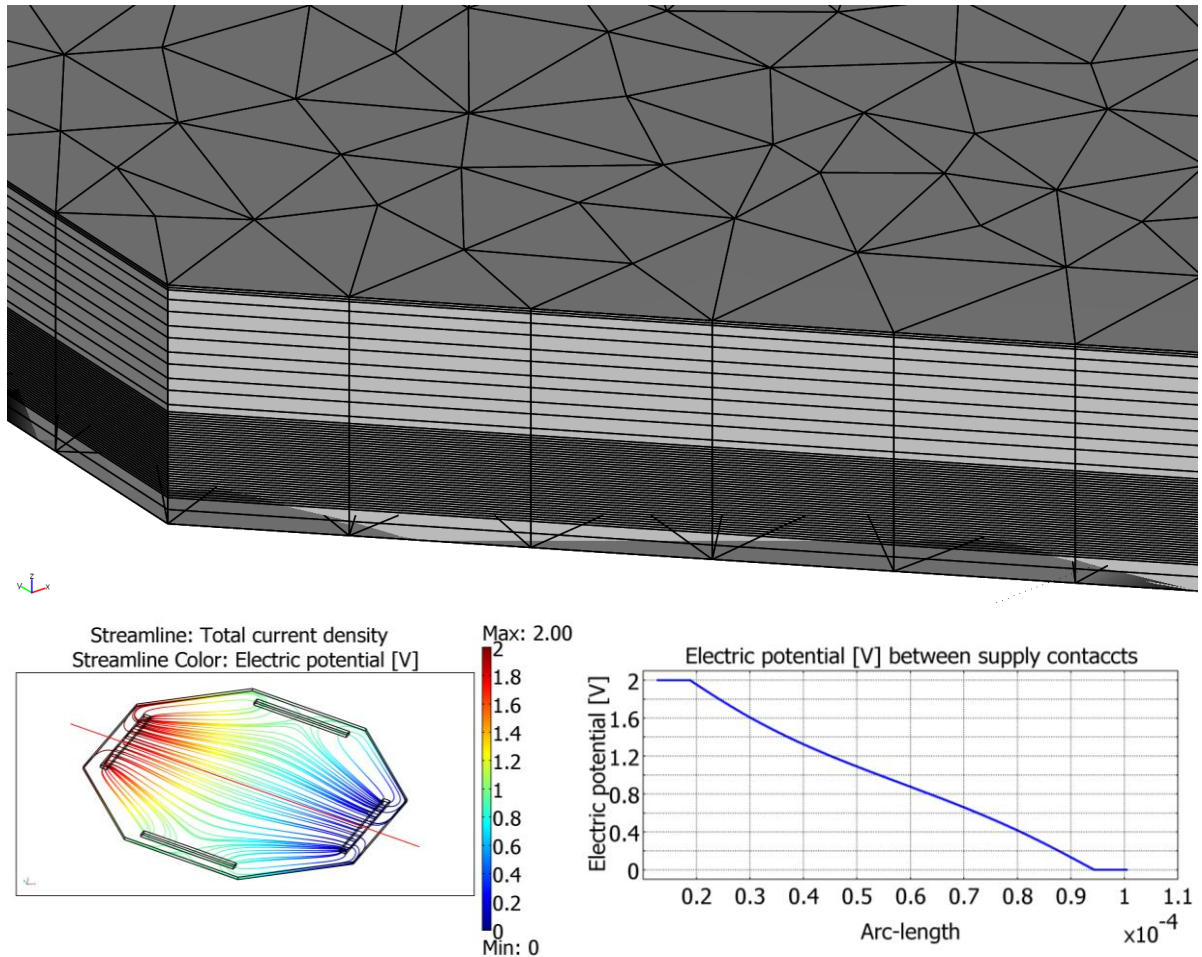

FIGURE A1: Finite element model of the junction field effect in Hall plates. The upper figure shows the meshing, which is triangular in (x,y)-plane and stratified in z-direction. The mesh layers are thinner near the bottom of the Hall tub where the junction field effect takes place and the pn-isolation interface varies according to the potential distribution. The lower figures show the current streamlines and potential distribution. The potential along the red line between both supply contacts is plotted in the right lower figure. The red line is at 45° to the x- and y-axes. [Figure in black & white]

## Appendix B: Conductivity tensor at mechanical stress and magnetic field

Most finite element codes need to specify conductivity instead of resistivity. Unfortunately, conductivity has larger expressions.

$$\kappa_{effective,xx}^{chip} = \rho^{-1}\left(1 - \frac{\pi_{11}+\pi_{12}+\pi_{44}}{2}\sigma_{xx}^{chip} - \frac{\pi_{11}+\pi_{12}-\pi_{44}}{2}\sigma_{yy}^{chip} - \pi_{12}\sigma_{zz}^{chip}\right) \quad (B1)$$

$$\kappa_{effective,yy}^{chip} = \rho^{-1}\left(1 - \frac{\pi_{11}+\pi_{12}-\pi_{44}}{2}\sigma_{xx}^{chip} - \frac{\pi_{11}+\pi_{12}+\pi_{44}}{2}\sigma_{yy}^{chip} - \pi_{12}\sigma_{zz}^{chip}\right) \quad (B2)$$

$$\kappa_{effective,zz}^{chip} = \rho^{-1}\left(1 - \pi_{12}\sigma_{xx}^{chip} - \pi_{12}\sigma_{yy}^{chip} - \pi_{11}\sigma_{zz}^{chip}\right) \quad (B3)$$

$$\kappa_{effective,yz}^{chip} = -\rho^{-1}\pi_{44}\sigma_{yz}^{chip} + \rho^{-2}C_H \times \quad (B4)$$

$$\times \begin{pmatrix} 1 - \frac{\pi_{11}+3\pi_{12}-\pi_{44}-P_{11}-P_{12}-P_{44}}{2}\sigma_{xx}^{chip} - \frac{\pi_{11}+3\pi_{12}+\pi_{44}-P_{11}-P_{12}+P_{44}}{2}\sigma_{yy}^{chip} - (\pi_{11}+\pi_{12}-P_{12})\sigma_{zz}^{chip} \\ (\pi_{11}-\pi_{12}+P_{11}-P_{12})\sigma_{xy}^{chip} \\ (\pi_{44}+P_{44})\sigma_{xz}^{chip} \end{pmatrix} \cdot \begin{pmatrix} B_x^{chip} \\ B_y^{chip} \\ B_z^{chip} \end{pmatrix}$$

$$\kappa_{effective,xz}^{chip} = -\rho^{-1}\pi_{44}\sigma_{xz}^{chip} - \rho^{-2}C_H \times \quad (B5)$$

$$\times \begin{pmatrix} (\pi_{11}-\pi_{12}+P_{11}-P_{12})\sigma_{xy}^{chip} \\ 1 - \frac{\pi_{11}+3\pi_{12}+\pi_{44}-P_{11}-P_{12}+P_{44}}{2}\sigma_{xx}^{chip} - \frac{\pi_{11}+3\pi_{12}-\pi_{44}-P_{11}-P_{12}-P_{44}}{2}\sigma_{yy}^{chip} - (\pi_{11}+\pi_{12}-P_{12})\sigma_{zz}^{chip} \\ (\pi_{44}+P_{44})\sigma_{yz}^{chip} \end{pmatrix} \cdot \begin{pmatrix} B_x^{chip} \\ B_y^{chip} \\ B_z^{chip} \end{pmatrix}$$

$$\kappa_{effective,xy}^{chip} = \rho^{-1}(\pi_{12}-\pi_{11})\sigma_{xy}^{chip} + \rho^{-2}C_H \begin{pmatrix} (\pi_{44}+P_{44})\sigma_{xz}^{chip} \\ (\pi_{44}+P_{44})\sigma_{yz}^{chip} \\ 1-(\pi_{11}+\pi_{12}-P_{12})(\sigma_{xx}^{chip}+\sigma_{yy}^{chip}) - (2\pi_{12}-P_{11})\sigma_{zz}^{chip} \end{pmatrix} \cdot \begin{pmatrix} B_x^{chip} \\ B_y^{chip} \\ B_z^{chip} \end{pmatrix} \quad (B6)$$

## Appendix C: Mechanical stress in a laminate

Following [27], the strain in a laminate is a linear function of thickness coordinate z (with z = 0 at the bottom of the laminate):

$$\varepsilon = \varepsilon_{bottom} + (\varepsilon_{top} - \varepsilon_{bottom})z/T \quad (C1)$$

For cylindrical bending there is no strain in width direction and therefore stress $\sigma$ and strain in the i-th layer (i = 1, 2) are linked via

$$\varepsilon = CTE_i(Temp - Temp_{stress-free}) + \sigma(1-v_i^2)/E_i \quad (C2)$$

With the coefficient of thermal expansion $CTE_i$, the homogeneous temperature *Temp* of the laminate, Poisson number $v_i$ and Young's modulus $E_i$. The layers are glued together without any stress at the temperature $Temp_{stress-free}$. The shearing force in *i*-th layer of the laminate is given by

$$Q_i = W \int_{z=z_{i-1}}^{z_i} \sigma\, dz \tag{C3}$$

With $z_0 = 0$, $z_2 = T$, and $z_1$ being the thickness of the bulk silicon, *i.e.* 739.7 µm. The bending moment in the laminate beam is

$$M_i = W \int_{z=z_{i-1}}^{z_i} z\sigma\, dz \tag{C4}$$

The balances of forces and moments give two equations

$$\sum_{i=1}^{2} Q_i = 0 \tag{C5}$$

$$\sum_{i=1}^{2} M_i = F_{applied}\left(L_o - L_i\right)/4 \tag{C6}$$

which can be solved for $\varepsilon_{bottom}$ and $\left(\varepsilon_{top} - \varepsilon_{bottom}\right)$. Inserting this into (C2) and solving for $\sigma$ gives the stress an all layers. The result is a long equation. For our wafer stripe bending experiment we use the silicon parameters in <100> direction, $E_1$ = 130.2 GPa, $v_1$ = 0.279.